\documentclass{aastex62}

\usepackage{epstopdf}
\usepackage{rotating}

\newcommand{\upp}

\begin{document}
\title{Giant X-ray and optical Bump in  GRBs: evidence for fall-back accretion model}

\correspondingauthor{He Gao}
\email{gaohe@bnu.edu.cn}

\author[0000-0001-6437-0869]{Litao Zhao}
\affil{Department of Astronomy ,
Beijing Normal University, Beijing, China;}

\author[0000-0002-3100-6558]{He Gao}
\affiliation{Department of Astronomy ,
Beijing Normal University, Beijing, China;}

\author[0000-0003-3440-1526]{WeiHua Lei}
\affiliation{Department of Astronomy, School of Physics, Huazhong University of Science and Technology, Wuhan, China}

\author{Lin Lan}
\affiliation{Department of Astronomy ,
Beijing Normal University, Beijing, China;}

\author[0000-0002-8708-0597]{Liangduan Liu}
\affiliation{Department of Astronomy ,
Beijing Normal University, Beijing, China;}

\begin{abstract}
The successful operation of dedicated detectors has brought us valuable information for understanding the central engine and the progenitor of gamma-ray bursts (GRBs).  For instance, the giant X-ray and optical bumps found in some long-duration GRBs (e.g. GRBs 121027A and 111209A) imply that some extended central engine activities, such as the late X-ray flares, are likely due to the fall-back of progenitor envelope materials. Here we systemically search for long GRBs that consist of a giant X-ray or optical bump from the Swift GRB sample, and eventually we find 19 new possible candidates. The fall-back accretion model could well interpret the X-ray and optical bump for all candidates within a reasonable parameter space. Six candidates showing simultaneous bump signatures in both X-ray and optical observations, which could be well fitted at the same time when scaling down the X-ray flux into optical by one order of magnitude, are consistent with the standard $F_{\nu}\propto\nu^{1/3}$ synchrotron spectrum.
The typical fall-back radius is distributed around $10^{10}\rm-10^{12}$ cm, which is consistent with the typical radius of a Wolf-Rayet star. The peak fall-back accretion rate is in the range of $\sim 10^{-11}-10^{-4}M_{\odot} \ \text{s} ^{-1}$ at time $\sim10^{2}- 10^{5}~\rm s$, which is relatively easy to fulfill as long as the progenitor's metallicity is not too high. Combined with the sample we found, future studies of the mass supply rate for the progenitors with different mass, metallicity, and angular momentum distribution would help us to better constrain the progenitor properties of long GRBs.
\end{abstract}


\section{Introduction} \label{sec:intro}

Increasing evidence suggests  that the long gamma-ray bursts (GRBs) are associated with the death of massive stars \citep{woosley93,paczynski98,macfadyenwoosley99,woosley06}. At the end of a massive star's life, electron-trapping and photon decomposition will trigger core collapse and form a hyperaccreting black hole (BH) or a rapidly spinning magnetar, which can launch a relativistic jet. The internal dissipation of the relativistic jet fuels the prompt emission, and the external shocks (especially the forward shock) due to jet-medium interaction contributes multiwavelength afterglow emission (see \cite{zhangbook} for a review).

In general, the end of the prompt emission phase means the cease of the central engine. However, the observations of Neil Gehrels Swift suggest that many GRBs have an extended central engine activity time, manifested through flares \citep{burrows05,zhang06,margutti11} and extended shallow plateaus \citep{troja07,liang07,zhao19,tang19} in the X-ray light curves following the MeV emission. It has long been proposed that some of these interesting signatures could help us to determine the central engines for particular GRBs \citep{dailu98,rees98,zhang01,zhang06,nousek06}. For instance, systematic analysis for the \emph{Swift} GRB X-ray afterglow shows that bursts with X-ray plateau features likely have rapidly spinning magnetars as their central engines \citep{liang07,zhao19,tang19}, especially when the X-ray plateau followed by a very steep decay. The steep decay is difficult to be interpret within the framework of a BH central engine, but is consistent within a magnetar engine picture, where the abrupt decay is interpreted as the collapse of a supramassive magnetar into a BH after the magnetar spins down \citep{troja07,lyons10,rowlinson10,rowlinson13,lu14,lu15,gao16a,depasquale16,ZhangQ16}. Recently, \cite{chen17} found one candidate, GRB 070110, that  showed a small X-ray bump following its internal plateau, and \cite{zhao20} found another three candidates in the Swift sample, i.e., GRBs 070802, 090111, and 120213A, whose X-ray afterglow light curves contain two plateaus, with the first one being an internal plateau. These particular cases provide further support to the magnetar central engine model.

For GRBs without shallow decay features, their most promising central engine should be the hyperaccreting BH system. In this scenario, the late X-ray flares could be interpreted with four different approaches: 1) part of the massive star envelope mass falling back onto the BH and reactivating the central engine \citep{kumar08a,kumar08b}; 2) late time features need not necessarily be related to late central engine activity, since they might be due to the late internal collisions or refreshed external collisions from early ejected shells \citep{rees98,sari00,gao15}; 3) late flares can arise from the interaction of a long-lived reverse shock (RS) with a stratified ejecta produced by a gradual and nonmonotonic shutdown of the central engine right after the initial ejection phase \citep{uhm07,genet07,hascoet17}; 4) an outflow of modest Lorentz factor is launched more or less simultaneously with the highly relativistic jet that produced the prompt gamma-ray emission, so that flares are produced when the slow moving outflow reaches its photosphere \citep{beniamini16}.

For the fallback accretion model, if the fallback accretion rate and fallback duration are large enough,  the giant X-ray and optical bump with rapid rising and $t^{-5/3}$ decaying feature are expected, which can hardly be interpreted with the latter three candidate models. Up to now, such giant X-ray bumps have been discovered in two GRBs, 121027A and 111209A, and both data could be well interpreted under the fall-back accretion model framework \citep{wu13,yu15,gao16b}.

Thanks to the successful operation of dedicated satellites and  ground-based detectors, many GRBs were detected with good quality X-ray and optical afterglow observations. In this work, we systematically search for long GRBs with giant X-ray or optical bump from the GRB sample. The data reduction method and the sample selection results are presented in section 2. In section 3, we described the fall-back accretion model and apply this model to the giant X-ray and optical bump observed in our selected sample. The conclusion and implications of our results are discussed in Section 4. Throughout the paper, the convention $Q=10^nQ_n$ is adopted in c.g.s. units.

\setlength{\tabcolsep}{1.2mm}
\begin{deluxetable*}{ccCclccccccc}[b!]
\tablecaption{ Features of the sample \label{table-1}}
\tablecolumns{6}
\tablenum{1}
\tablewidth{0pt}
\tablehead{
\colhead{GRB} &
\colhead{$\log_{10}(t_{\rm start})$\tablenotemark{a}} &
\colhead{$\log_{10}(t_{\rm end})$\tablenotemark{a}} &
\colhead{Instruments}&
\colhead {$T_{90}$} &
\colhead {$\Gamma_{\gamma}$} &
\colhead {$S_{\gamma}$} &
\colhead{$\rm{Reference}^{b}$}&
\colhead{$z$}&
\colhead{$\rm{Reference}^{c}$}&
\colhead{$\rm{Reference}^{d}$}&
\colhead{ Group}\\
\colhead{} &\colhead{} & \colhead{s} &\colhead{s} & \colhead{s}  & \colhead{}  & \colhead{$10^{-7} \rm erg\ cm^{-2} $ }&\colhead{}&
}
\startdata
970508 & 4.81 & 6.11&BeppoSAX& $20$&3.7$\pm$0.1&$34\pm7$&A02 A08&0.835&Q13& D18&Silver\\
020903 &4.92  & 6.01&HETEC2&$10$&$3.46^{+0.44}_{-0.82}$&$1.6\pm0.4$&A08 S04a&0.25& Q13&D18 S04b&Silver\\
050814 & 2.91 & 4.17&Swift&$65^{+40}_{-20}$& $1.8\pm0.17$&$21.7\pm3.6$ &T05 &5.3 &J06a &J06b&Silver\\
051016B & 2.53 & 4.13&Swift&4$\pm$0.47&2.38$\pm$0.23&1.7$\pm$0.2&B05&0.9364&S05&X10&Silver\\
060206&3.41&4.51&Swift&$7\pm2$&$1.06\pm0.34$&$8.4\pm0.4$&P06a&4.048&P06b&D18 K10&Gold\\
060906 & 3.75 & 4.55&Swift&43.6$\pm$1&  2.02$\pm$0.11 &22.1$\pm$1.4&S06a S06b&3.6856& F09b&L13&Gold\\
070103 & 2.32 & 3.14&Swift&19$\pm$1&2$\pm$0.2&3.4$\pm$0.5&B07&2.6208&K12&...&Silver\\
071010A & 4.51 & 5.73&Swift&6$\pm$1&2.33$\pm$0.37&2$\pm$0.4&K07&0.98& P07&L13&Gold\\
081028 & 3.95 & 4.85&Swift& 260$\pm$40 & 1.25$\pm$0.38 & 37$\pm$2 &B08a& 3.038 & B08b&M08 R08&Silver\\
081029&3.51&4.41&Swift&$270\pm45$&$1.43\pm0.18$&$21\pm2$&C08b&3.847& C08a&D18&Gold\\
090205 & 2.42& 3.25&Swift&8.8$\pm$1.8&2.15$\pm$0.23&1.9$\pm$0.3&C09&4.6497&F09a&D10&Silver\\
100418A&3.72&6.51&Swift&$7\pm1$&$2.16\pm0.25$&$3.4\pm0.5$&U10&0.624&C10b& D18 L15&Gold\\
100901A&4.01&5.12&Swift&$439\pm33$&$1.52\pm0.21$&$21\pm3$&S10&1.315&C10a&D18 L15&Gold\\
110715A & 4.22 & 5.41&Swift&13$\pm$4&1.25$\pm$0.12&118$\pm$2 &U11&0.82&P11b&S13&Silver\\
111209A&3.31&3.92&Swift&1400&1.48$\pm$0.03&360$\pm$10&P11a&0.677&V11&...&Silver\\
120118B & 2.61& 4.92&Swift&23.26$\pm$4.02&2.08$\pm$0.11&18$\pm$1&S12&2.943&M13&...&Silver\\
121027A&3.06&4.3&Swift&62.6$\pm$4.8&1.82$\pm$0.09&20$\pm$1&B12&1.77&T12&...&Silver\\
130831A&2.67&3.71&Swift&$32.5\pm2.5$&$1.93\pm0.05$&$65\pm2$&B13&0.4791&C13&D18 K19a&Silver\\
140515A & 2.75 & 4.22&Swift&23.4$\pm$2.1&0.98$\pm$0.64&5.9$\pm$0.6&S14&6.32&C14&M15&Silver\\
161129A& 2.23 & 2.85&Swift&35.53$\pm$2.09&1.57$\pm$0.06&36$\pm$1&B16&0.645&C16&K16a K16b Y16&Silver\\
190829A& 2.72& 3.93&Swift&58.2$\pm$8.9&2.56$\pm$0.21&64$\pm$7 &L19&0.078&D19&C19 K19b V19&Silver\\
\enddata
\tablenotetext{a}{$t_{0}$ and $t_{1}$ are  beginning  time  and end time of the  giant X-ray or optical bump, respectively.  }
\tablenotetext{b}{ The references of GRB prompt phase observations  for our sample. }
\tablenotetext{c}{ The references of  GRB redshift for our sample. }
\tablenotetext{d}{ The references of  optical afterglow data  for our sample. }
\tablecomments{References: (A02)\cite{amati02};(A08)\cite{amati08};(B05)\cite{2005GCN..4104....1B};(B07)\cite{ 2007GCN..5991....1B}; (B08a)\cite{2008GCN..8428....1B};  (B08b)\cite{2008GCN..8434....1B}; (B12)\cite{2012GCN.13910....1B}; (B13)\cite{2013GCN.15155....1B}; (B16)\cite{ 2016GCN.20220....1B}; (C08a)\cite{2008GCN..8448....1C}; (C08b)\cite{2008GCN..8447....1C};  (C09)\cite{2009GCN..8886....1C}; (C10a)\cite{2010GCN.11164....1C}; (C10b)\cite{2010GCN.10624....1C}; (C13)\cite{2013GCN.15144....1C}; (C14)\cite{2014GCN.16269....1C};  (C16)\cite{2016GCN.20245....1C}; (C19) \cite{2019GCN.25569....1C}; (D10)\cite{davanzo10}; (D18)\cite{deugarte18};  (D19)\cite{2019GCN.25552....1D};  (F09a)\cite{2009GCN..8892....1F}; (F09b)\cite{fynbol09};(J06a)\cite{jakobsson06a}; (J06b)\cite{jakobsson06b}; (K07)\cite{2007GCN..6868....1K};(K10)\cite{kann10}; (K12)\cite{kruhler12}; (K16a)\cite{2016GCN.20217....1K}; (K16b)\cite{2016GCN.20218....1K};  (K19a)\cite{klose19}; (K19b)\cite{2019GCN.25560....1K}; (L13)\cite{liang13}; (L15)\cite{laskar15};(L19)\cite{2019GCN.25579....1L}; (M08)\cite{2008GCN..8499....1M}; (M13)\cite{2013GCN.14225....1M}; (M15)\cite{melandri15}; (P06a)\cite{2006GCN..4697....1P};(P06b)\cite{2006GCN..4701....1P}; (P07)\cite{2007GCN..6864....1P};  (P11a)\cite{2011GCN.12640....1P};(P11b)\cite{2011GCN.12164....1P};  (Q13)\cite{qin13};  (R08)\cite{2008GCN..8455....1R}; (S04a)\cite{sakamoto04}; (S04b)\cite{soderberg04}; (S05)\cite{2005GCN..4186....1S};(S06a)\cite{2006GCN..5534....1S};(S06b)\cite{2006GCN..5538....1S}; (S10)\cite{2010GCN.11169....1S}; (S12)\cite{2012GCN.12873....1S};(S13)\cite{sanchezram13}; (S14)\cite{2014GCN.16284....1S}; (T05)\cite{2005GCN..3803....1T}; (T12)\cite{2012GCN.13929....1T}; (U10)\cite{2010GCN.10615....1U}; (U11)\cite{ 2011GCN.12160....1U}; (V11)\cite{2011GCN.12648....1V}; (V19)\cite{2019GCN.25565....1V}; (X10)\cite{xu10};(Y16)\cite{2016GCN.20223....1Y}. }
\end{deluxetable*}

\section{Data Reduction and Sample Selection}
For the purpose of this work, we systemically search for long GRBs consisting of a giant X-ray or optical bump from the GRB sample (detected between 1997 January and 2019 October).  The XRT light-curve data were downloaded from the Swift/XRT team website\footnote{\url{http://www.swift.ac.uk/xrt\_ curves/}}\citep{belokurov09}, and processed through HEASOFT (v6.12) software. The optical afterglow data were searched from published papers. 
Compared with typical flares, the duration of the giant bump should be relatively longer. \cite{yi16} have analyzed all significant X-ray flares from the GRBs observed by Swift from 2005 April to 2015 March, and obtained an empirical relationship:
\begin{eqnarray}
\log_{10}T_{\rm{dur}}=(-0.35\pm0.11)+(1.12\pm0.04)\times \log_{10}(T_{\rm{peak}}).
\label{Tp-Td}
 \end{eqnarray}
where $T_{\rm{peak}}$ is the peak time of the flare, and $T_{\rm{dur}}=T_{\rm end}-T_{\rm start}$ is the flare duration, with $T_{\rm start}$ and $T_{\rm end}$ being the start time and end time of flares. For each flare, $T_{\rm{peak}}$ and $T_{\rm{dur}}$ could be easily obtained by fitting the light curve with a smooth broken power-law function (representing the flare) superposing on a simple power-law function (representing the underlying continuum component), see details in \cite{yi16}. Here we define the flares of $ 2 \sigma$ deviate from the $T_{\rm{peak}}-T_{\rm{dur}}$ empirical relationship (Eq. \ref{Tp-Td}) to the longer duration trend as a giant bump.

\begin{figure}[hbt]
         \figurenum{1}
	\centering
	\includegraphics[width=12cm,height=8cm]{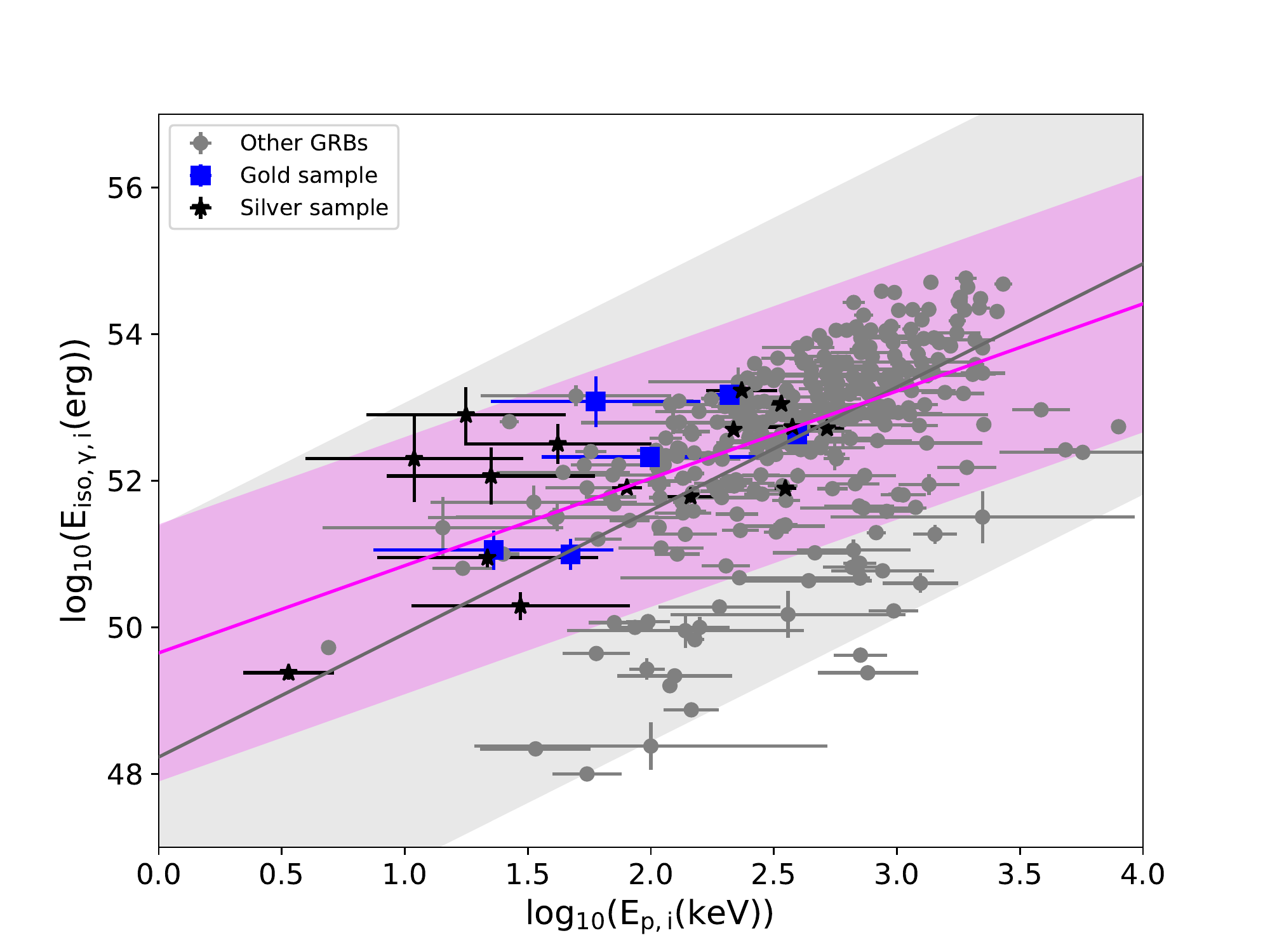}
	\caption{The Amati relation ($E_{\rm iso,\gamma,i}-E_{p,i}$)  in our sample, compared with other GRBs that have redshift measurements and are detected up to 2019 October.  We apply the regression model to GRBs in our sample and other GRBs, yielded $ \log_{10}(E_{\rm iso,\gamma,i})=(50.07\pm0.55)+(1.05\pm0.28)\log_{10}(E_{p,i})$ and $ \log_{10}(E_{\rm iso,\gamma,i})=(48.75\pm0.33)+(1.48\pm0.13)\log_{10}(E_{p,i} )$, respectively. The violet line mark the best fitting result for  our sample and the light  violet shadowed region shows the intrinsic scatter to the population 3$\sigma$. The grey line marks the best fitting result for other GRBs and the light gray shadowed region shows the intrinsic scatter to the population 3$\sigma$.}
	\label{Amati}
\end{figure}

Besides GRB 121027A and GRB 111209A, we find another 19 GRB candidates containing a giant X-ray or optical bump signature that could meet our selection criteria. We collect the names and optical data references of these candidates in Table \ref{table-1}, together with their basic observational properties, such as the prompt emission duration $T_{90}$, the detected gamma-ray fluences  $S_{\gamma}$, the power-law photon index of prompt emission spectrum  $\Gamma_ {\gamma}$, the beginning time $t_{\rm start}$ and ending time $t_{\rm end}$ of the giant X-ray or optical bump, and the redshift $z$.  For the whole sample, the $T_{90}$, $S_{\gamma}$, $\Gamma_ {\gamma}$ and $z$ are distributed in the ranges of  $4-1400$ s, $1.6 \times10^{-7}-3.36 \times 10^{-5}~ \rm erg\ cm^{-2} $, $0.98-3.7$, and $0.078-6.32$, respectively.  The beginning time of the giant X-ray or optical bump is distributed in the range of $169 - 8.32\times10^{4}$ s,  and the duration of the bump is in an extended range of $538-3.23\times10^{6}$ s.  Here we divide the new candidates into two samples: the gold sample that contains simultaneous bump signatures in both X-ray and optical afterglow data (6/19), and the silver sample that shows giant X-ray but without giant optical bump or vice versa (13/19).  It is worth noting that GRB 130831A contains detections from both X-ray and optical band, but only optical bands has enough data when the bump signature emerges, and it is thus assigned to the silver sample. Due to the lack of simultaneous bump signatures, we put GRB 121027A and GRB 111209A in the silver sample.
 Figure \ref{Amati} shows the Amati relation ($E_{\rm iso,\gamma,i}-E_{p,i}$)\citep{amati02} for GRBs in our sample and other GRBs that have redshift measurements and were detected up to 2019 October \citep{minaev20}, where $E_{\rm iso,\gamma,i}$ is isotropic-equivalent radiation energies of the  GRB prompt phase in the $1-10^{4}$ keV band and $E_{p,i}$ is the peak energy of the $\nu F_{\nu}$ spectrum ($E_{p}$) in the burst rest frame, i.e., $E_{p,i}=E_{p}(1+z)$. It is interesting that bursts in our sample tend to have larger $E_{\rm iso,\gamma,i}$ for given $E_{p,i}$ values, which is consistent with the interpretation that these bursts may have more active fall-back accretion rates.

\section{Fall-back Accretion Model Application} \label{sec:model}
\subsection{Model Description}

In this paper, we intend to use the fall-back accretion model that has been described in \cite{wu13} to interpret giant X-ray and optical bumps in our selected sample. The physical picture is as follows: for our selected GRBs, their progenitor stars have a core-envelope structure, as is common in stellar models. At the end of the star's life, the bulk of the mass in the core part collapses into a rapidly spinning BH, and the rest mass forms a surrounding accretion disk. A relativistic jet is launched by the hyperaccreting BH system and it successfully penetrates the envelope to power the GRB prompt $\gamma$-ray and broadband afterglow emission. During the penetration, parts of the jet energy are transferred into the envelope, which might help the supernova to explode. The bounding shock responsible for the associated supernova would transfer kinetic energy to the envelope materials, so that most envelope materials would be ejected but with a small portion falling back onto the BH \citep{kumar08a,kumar08b}. The fall-back of the envelope materials may form a new accretion disk, powering a new relativistic jet through the Blandford-Znajek (BZ) mechanism \citep{blandfordznajek77,Lee00,Li00,lei05,lei13} or neutrino-annihilations mechanism\citep{popham99,narayan01,dimatteo12,janiuk04, gu06, chen07,liu07, liu15, lei09,lei17, xie16}.  In general cases, a BZ jet is more powerful than a neutrino-annihilation jet, which is more likely accounts for the central engine activities \citep{kawanaka13, lei17,xie17, Lloyd18}. Especially during the fall-back accretion stage, the typical accretion rate is far below the igniting accretion rate\footnote{Accretion in a fall-back disk can occur via two distinct modes. For high accretion rates, the disks are dense and hot enough in the inner regions to cool via neutrino losses. However, for lower fall-back rates and/or at larger radii, the accretion is radiatively inefficient and  has an advection-dominated accretion flow.  The accretion rate at the transition that the inner disk from being neutrino dominated to advection dominated is defined as igniting the accretion rate. As discussed in \cite{lei17}, the igniting accretion rate would be $0.07 M_\sun /s$ for a nonspinning BH and $0.01 M_\sun  s^{-1}$ for a fast spinning BH.}, the neutrino-annihilation power cannot explain the late-time X-ray activities in GRBs. Finally, a part of the jet energy would undergo internal dissipation and generate the observed giant X-ray and optical bump.

According to some analytical and numerical calculations, the evolution of the fall-back accretion rate can be described by a broken power-law function of time \citep{chevalier89,macfadyen01,zhangwei08,Dailiu12}
\begin{eqnarray}
\dot{M}_{acc}=\dot{M}_{\mathrm{p}}\left[\frac{1}{2}\left(\frac{t-t_{0}}{t_{\mathrm{p}}-t_{0}}\right)^{-s/2 }+\frac{1}{2}\left(\frac{t-t_{0}}{t_{p}-t_{0}}\right)^{-5s/3}\right]^{-1 / s},
\end{eqnarray}
where $t_{0}$ is the beginning time of the fall-back accretion in the cosmologically local frame, $t_{p}$ and $\dot{M}_{p}$ are the peak time and peak rate of fall-back accretion, and $s$ describes sharpness of the peak.

The  BZ power from a kerr BH with a mass $M_{\bullet}$ and angular momentum $J_{\bullet}$ could be estimated as \citep{Lee00,Li00,wang02,lei05,lei13,lei17,mckinney05,Leizhang13,chen17,liu17,Lloyd18}
 \begin{eqnarray}
 L_{\rm BZ}=1.7\times10^{50} {\rm erg\ s^{-1}} a_{\bullet}^{2}\left(\frac{M_{\bullet}}{M_{\odot}} \right)^{2} B_{\bullet, 15}^{2}F(a_{\bullet}),
 \end{eqnarray}
where $a_{\bullet}=J_{\bullet}c/GM_{\bullet}^{2}$ is dimensionless BH spin parameter, $B_{\bullet, 15}$ is strength of the magnetic field near the BH horizon in units of $10^{15}$ G, and
\begin{eqnarray}
F(a_{\bullet})=\left[\frac{1+q^{2}}{q^{2}}\right]\left[\left(q+\frac{1}{q}\right) \arctan q-1\right] ,
\end{eqnarray}
with $q= a_{\bullet}/(1+\sqrt{1-a_{\bullet}^{2}})$.

Accretion disk is essential for maintaining a strong magnetic field. If there is no accretion disk
magnetic pressure,  the magnetic field near the BH horizon will disappear quickly. We  estimate the value of  $B_{\bullet}$  by balancing the magnetic pressure on the BH horizon and ram pressure of the accretion flow at the inner edge of the accretion disk (\cite{modersiki97})

\begin{eqnarray}
\frac{B_{\bullet}^{2}}{8 \pi}=P_{\mathrm{ram}} \sim \rho c^{2} \sim \frac{\dot{M}_{acc} c}{4 \pi r_{h}^{2}},
\end{eqnarray}
where $r_{h}=(1+\sqrt{\left(1-a_{\bullet}^{2}\right)}) r_{g}$ is radius of BH horizon and $r_{g}=GM_{\bullet}/c^{2}$.  Therefore, the BZ power can be rewritten as
\begin{eqnarray}
L_{\rm BZ}=9.3 \times 10^{53} \frac{a_{\bullet}^{2} F\left(a_{\bullet}\right)}{(1+\sqrt{1-a_{\bullet}^{2}})^{2}}\frac{\dot{M}_{\rm acc}}{ (M_{\odot}\ s^{-1} )}~ \rm{erg} \ \mathrm{s}^{-1}.
\end{eqnarray}

We introduce the  X-ray and optical radiation efficiency $\eta_{(X,O)}$ and jet beaming factor $f_{b}=1-\rm cos\theta_{j}$ ($\theta_{j}$ is the jet opening angle) to connect the observed X-ray and optical luminosity $L_{(X,O)}$ and the BZ power $L_{\rm BZ}$ by
  \begin{eqnarray}
\eta_{(X,O)}L_{\rm BZ}=f_{b}L_{(X,O)}.
 \end{eqnarray}

Note that the BZ process extracts rotational energy and angular momentum from the BH, but the accretion process brings disk energy and angular momentum into the BH.  According to the conservation of energy and angular momentum, the evolution equation of BH under two processes are given as follows (\cite{wang02}):
\begin{eqnarray}
\frac{dM_{\bullet}c^{2}}{dt}=E_{\rm i n} \dot{M}_{\rm acc} c^{2}-L_{\rm BZ},
\label{eq-2}
\end{eqnarray}
and
\begin{eqnarray}
\frac{dJ_{\bullet}}{dt}=J_{\rm i n} \dot{M}_{\rm acc}-T_{\rm BZ}.
\label{eq-3}
\end{eqnarray}
$T_{\rm BZ}$ is the torque applied to BH by BZ process, which could be estimated by \citep{Li00,Leizhang13,lei17}
\begin{eqnarray}
T_{\rm BZ} =\frac{2L_{\rm BZ}}{\Omega_{H}} ,
\end{eqnarray}
where
\begin{eqnarray}
\Omega_{H}=\frac{c^{3}}{G M_{\bullet}} \frac{a_{\bullet}}{2(1+\sqrt{1-a_{\bullet}^{2}})},
\end{eqnarray}
is the angular velocity of BH horizon.

The evolution of dimensionless BH spin parameter $a_{\bullet}$  could   be derived from Equations \ref{eq-2} and \ref{eq-3},
\begin{eqnarray}
\frac{d a_{\bullet}}{d t}=\frac{\left(\dot{M}_{\rm acc} J_{\mathrm{\rm in}}-T_{\rm BZ}\right) c}{G M_{\bullet}^{2}}-\frac{2 a_{\bullet} \left(\dot{M}_{\rm acc} c^{2} E_{\mathrm{in}}-L_{\rm BZ}\right)}{M_{\bullet} c^{2}}.
\label{eq-4}
\end{eqnarray}
Here $E_{\rm in}$  and $J_{\rm in}$  are  represent  the specific energy and angular momentum  at the radius of the innermost inner edge of accretion disk $R_{\rm in}$, respectively, which are defined as \citep{novikov73}:
\begin{eqnarray}
 J_{\mathrm{in}} =\frac{G M_{\bullet} }{c} \frac{2(3 \sqrt{R_{\mathrm{in}}}-2 a_{\bullet})}{\sqrt{3R_{\rm in}} },
 \end{eqnarray}
 and
 \begin{eqnarray}
 E_{\mathrm{in}}=\frac{4 \sqrt{R_{\mathrm{in}}}-3 a_{\bullet}}{\sqrt{3} R_{\mathrm{in}}} ,
\end{eqnarray}
 where $R_{\rm in}$   can be obtained from \citep{{bardeen72}}
 \begin{eqnarray}
R_{\rm in}=\frac{r_{\rm in}}{r_{g}}=3+Z_{2}-\left[(3-Z_{1})(3+Z_{1}+2Z_{2})\right]^{1/2},
\end{eqnarray}
where $Z_{1}\equiv 1+(1-a_{\bullet}^{2})^{1/3}\left[(1+a_{\bullet})^{1/3}+(1-a_{\bullet})^{1/3}\right] $ and $Z_{2}\equiv(3a_{\bullet}^{2}+Z_{1}^{2})^{1/2}$ for $0\le a_{\bullet} \le 1$.

\subsection{Model Application to Selected GRBs}{\label{mcmcsimulation}}

In the following, we apply the fallback accretion model to fit  giant X-ray and optical bumps in our selected sample. Here we adopt the beginning  time and ending time of the giant X-ray or optical bump ($t_{\rm start},t_{\rm end}$) as the beginning time and ending time of the fall-back accretion. Since the initial BH mass $M_{0}$ hardly affects the BZ power, here we adopt $M_{0}=3~M_{\odot}$ as a fixed value for all GRBs in our sample. Considering that a BH may be spun up by accretion or spun down by the BZ mechanism, the BH spin will reach an equilibrium value $a_{\rm eq} \sim 0.87$ \citep{lei17}. For simplicity, we adopt $a_{0}=0.9$ as  a fixed value for all GRBs in our sample.
In our calculation, we take $\eta_{X}=0.01$ and $f_{b}=0.01$. For the gold sample, we simultaneously fit their X-ray and optical flux with $\eta_O=\xi\eta_X$. We take the dimensionless fallback accretion peak $\dot{m}_{p}$ ($\dot{m}_{p}=\dot{M}_{p}/M_{\odot}\ s^{-1} $), the peak sharpness of the fallback accretion $s$, the peak time of fallback accretion $t_{p}$ and  $\xi$ as our free parameters.  We then use the Markov chain Monte Carlo (MCMC) method to fit the data. In our fitting, we use a Python module emcee \citep{formanmackey13} to get best-fit values and uncertainties of free parameters. The allowed range of the four free parameters are set as $\log_{10}(\dot{m}_{p})\equiv[-15,0]$, $s\equiv[0,10]$, $t_{p}\equiv[t_{\rm start},t_{\rm end}]$, and $ \log_{10}(\xi)\equiv[-5,0]$.

\begin{figure}[hbt]
	\centering
	\figurenum{2}
         \includegraphics[width=5.86cm,height=6.5cm]{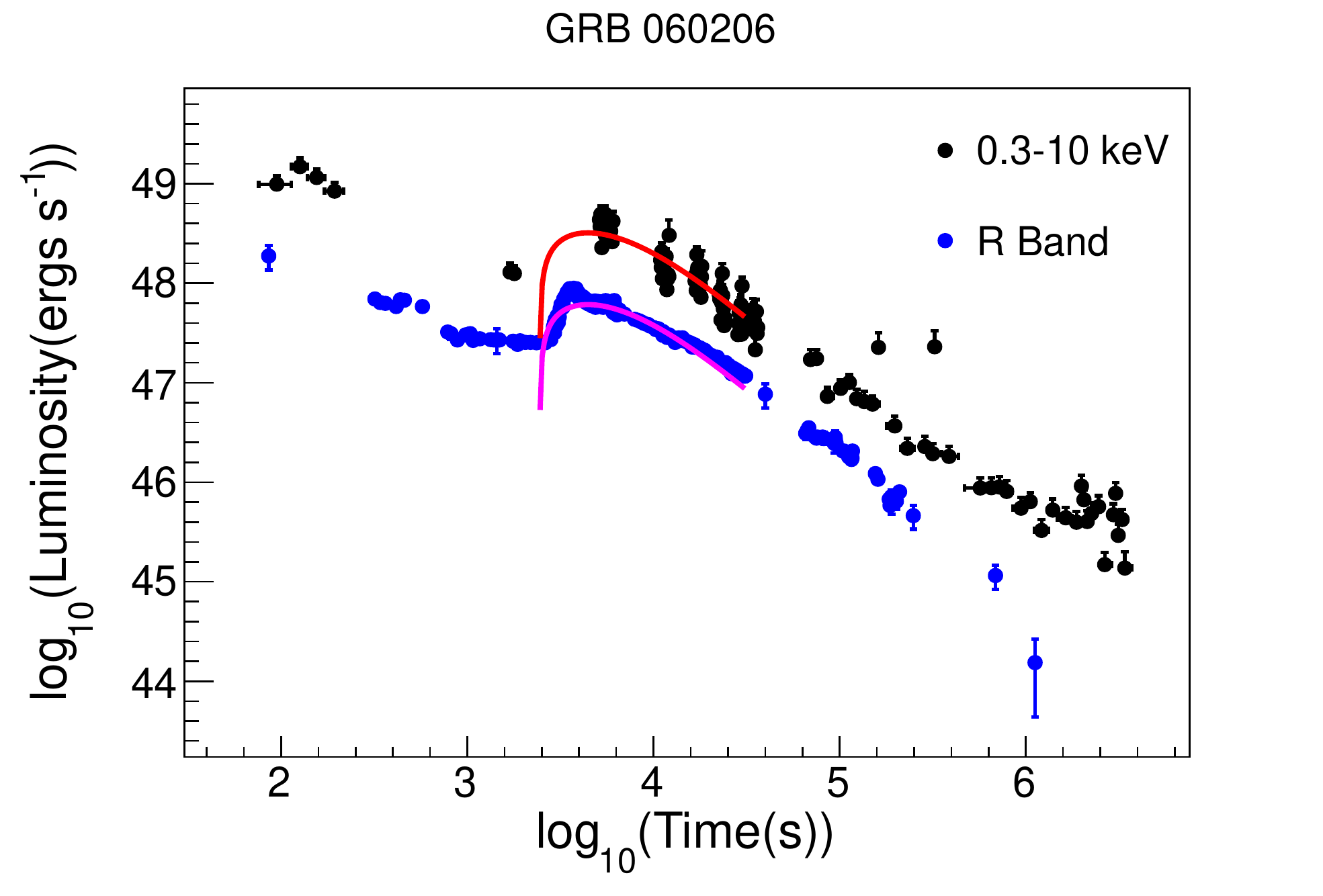}
         \includegraphics[width=5.86cm,height=6.5cm]{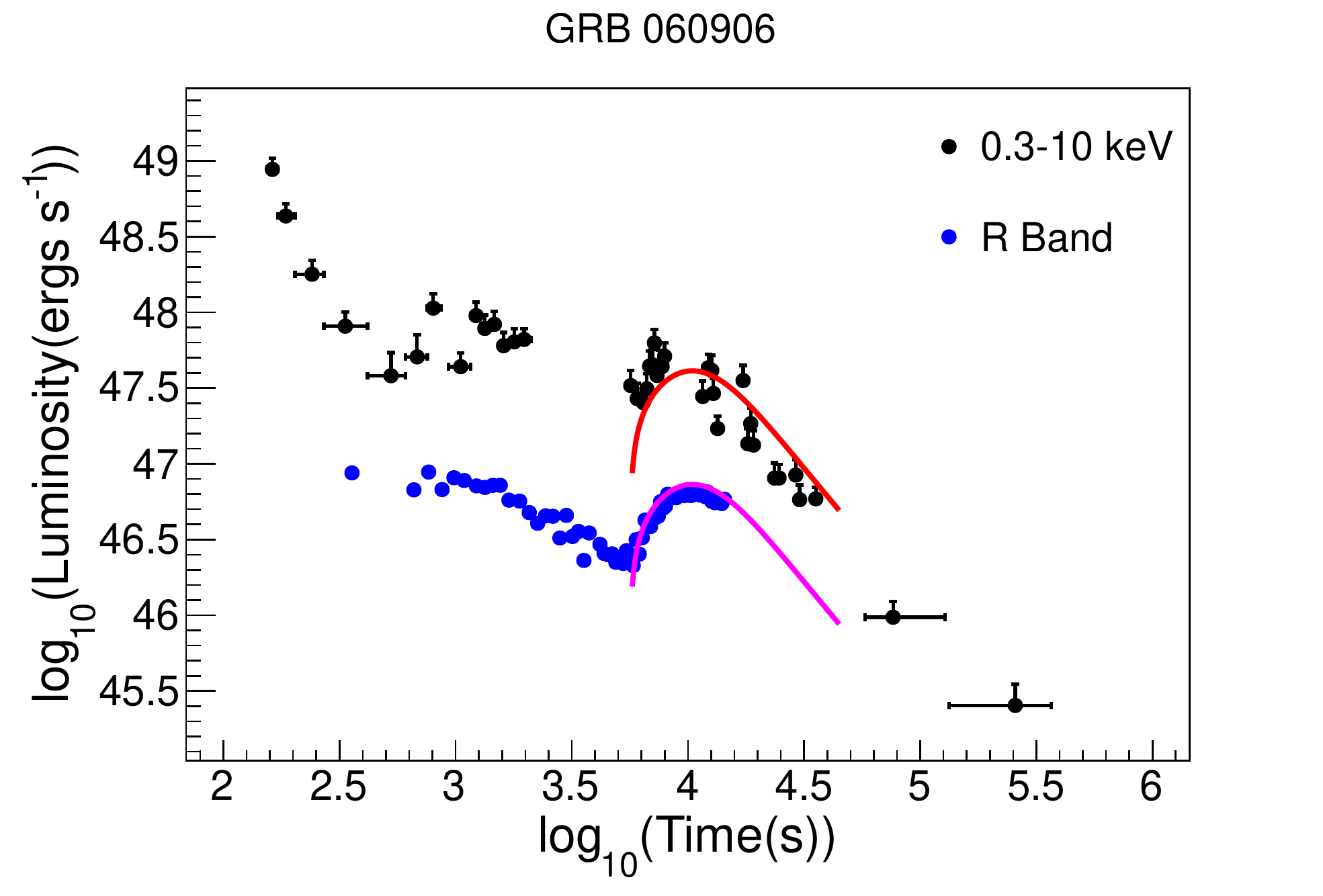}
         \includegraphics[width=5.86cm,height=6.5cm]{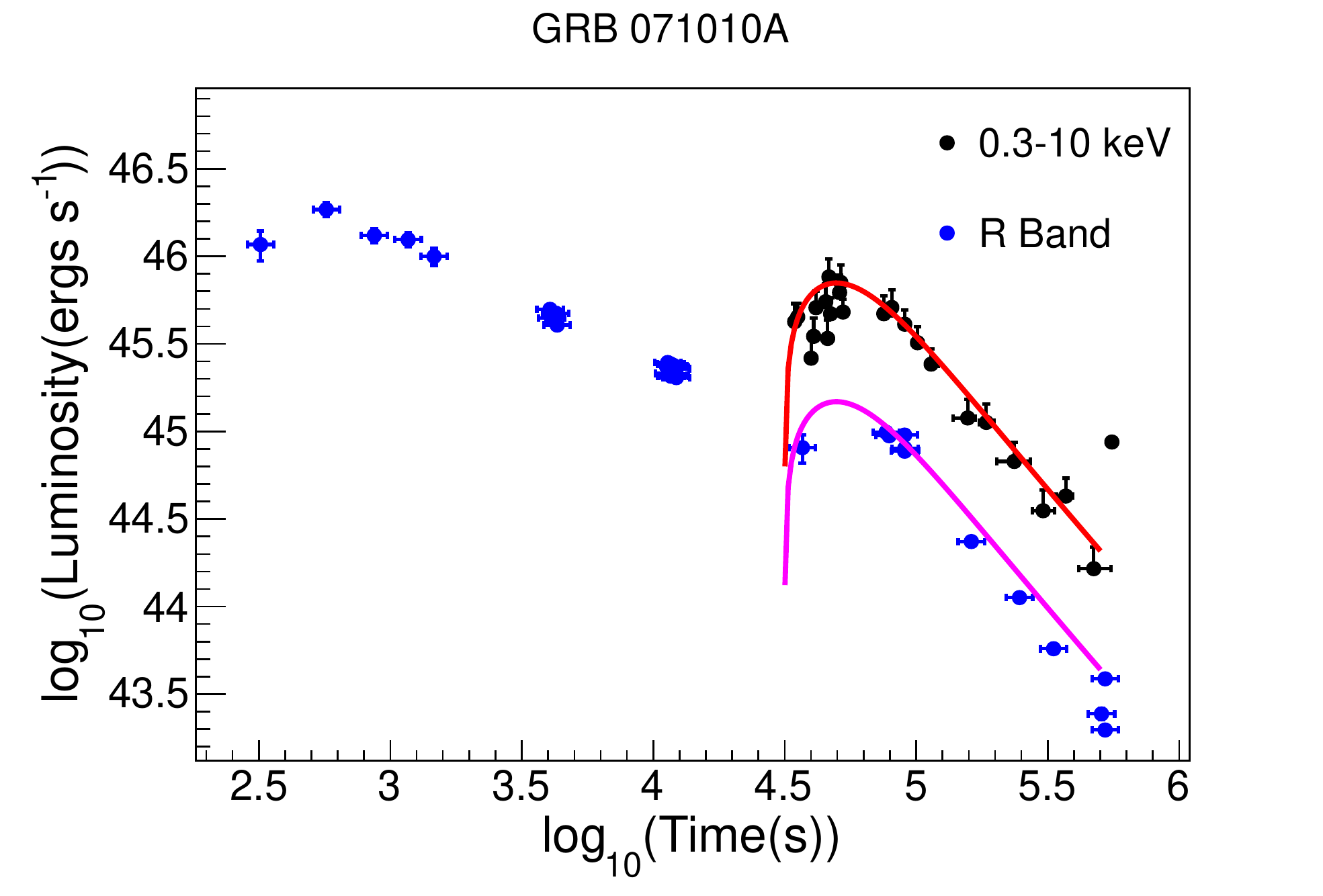}
         \includegraphics[width=5.86cm,height=6.5cm]{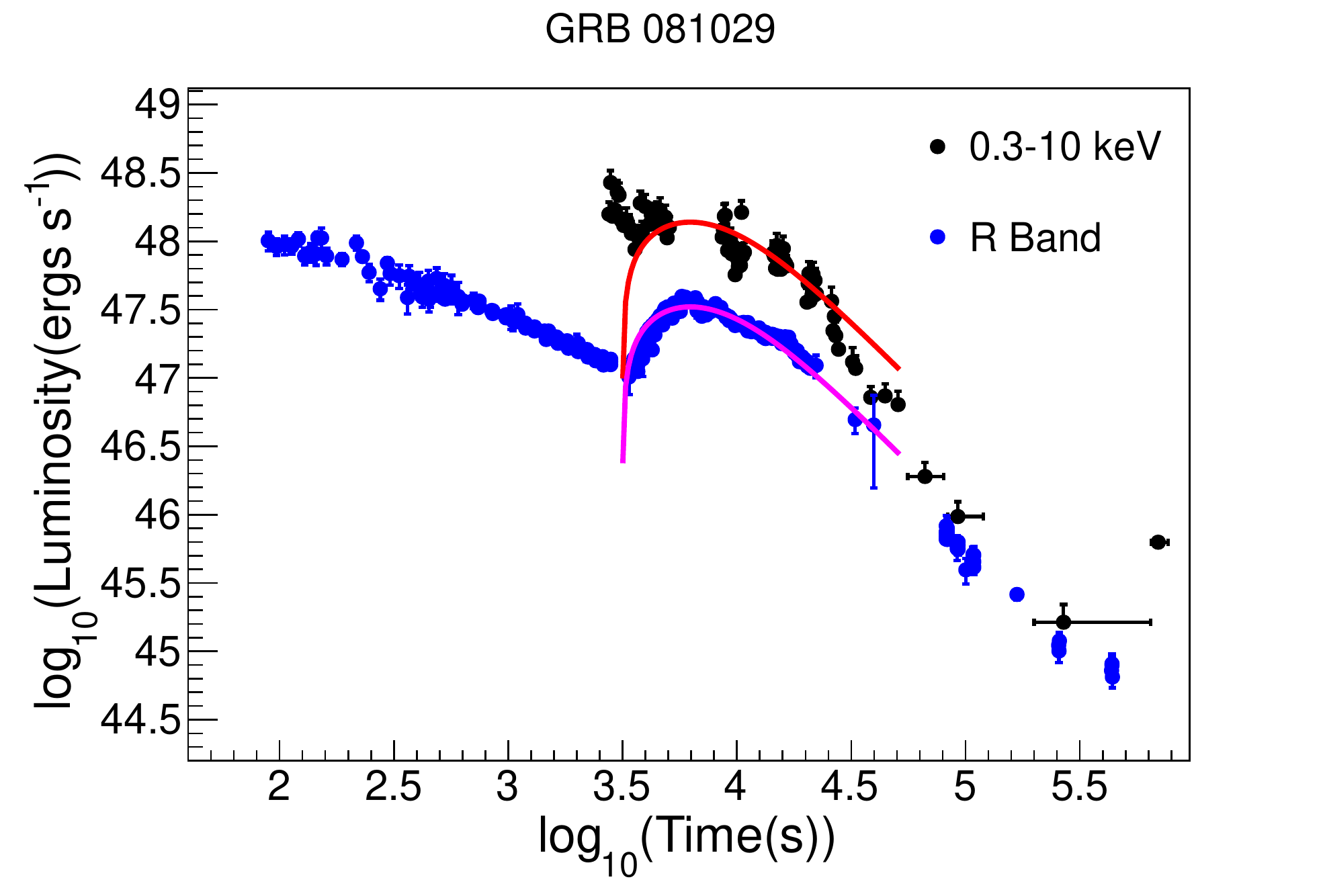}
         \includegraphics[width=5.86cm,height=6.5cm]{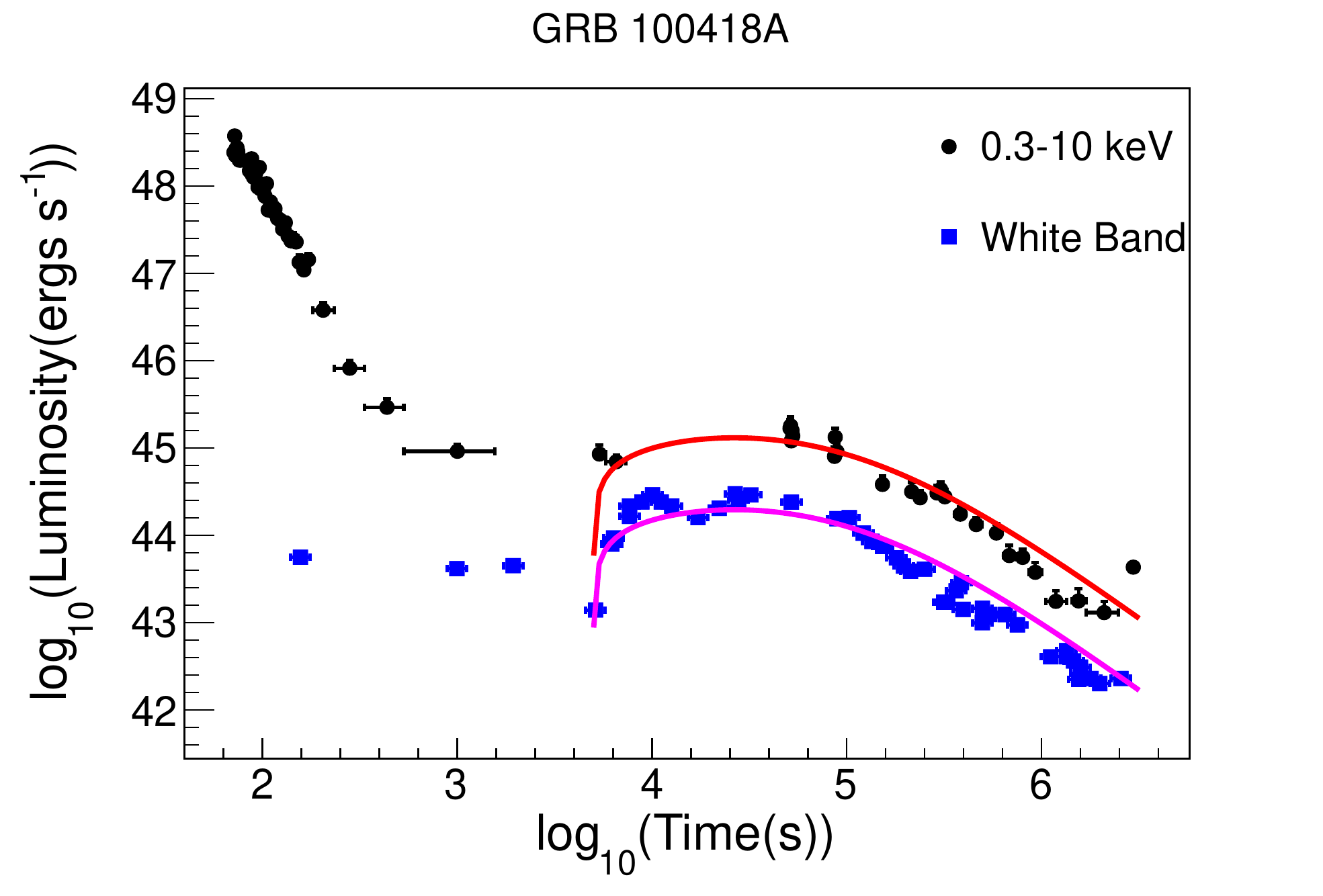}
         \includegraphics[width=5.86cm,height=6.5cm]{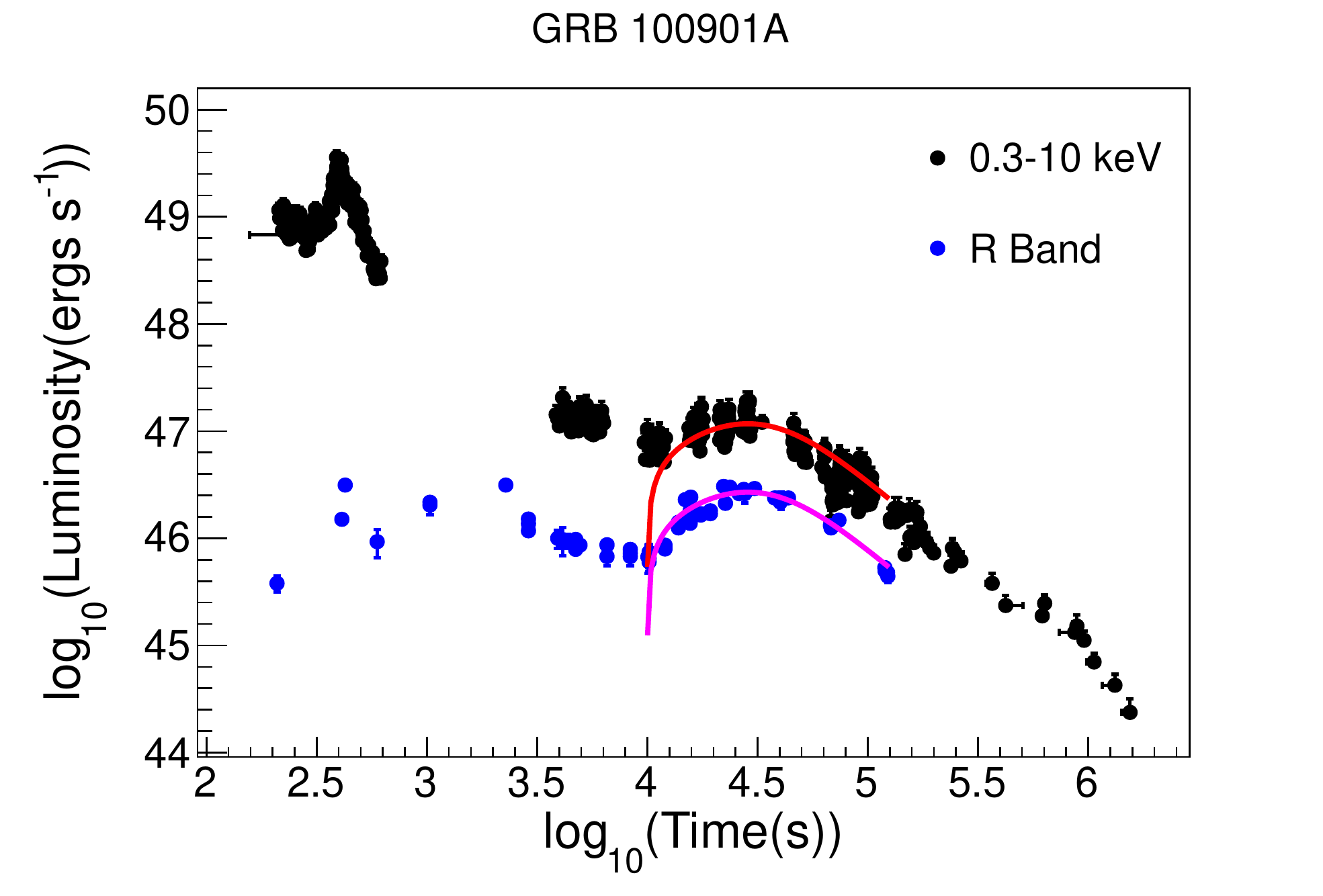}
         \caption{Modeling results for the giant X-ray and optical bumps  of the gold  sample. The red solid  lines and pink  solid  lines  presents  theoretical light curve  produced by our model within the the Swift-XRT energy band ($0.3-10$ keV) and optical band, respectively. } 
	\label{fittingresult-1}
\end{figure}

\begin{figure}[hbt]
       \figurenum{3}
	   \centering
	   \includegraphics[width=0.32\textwidth,height=6.5cm]{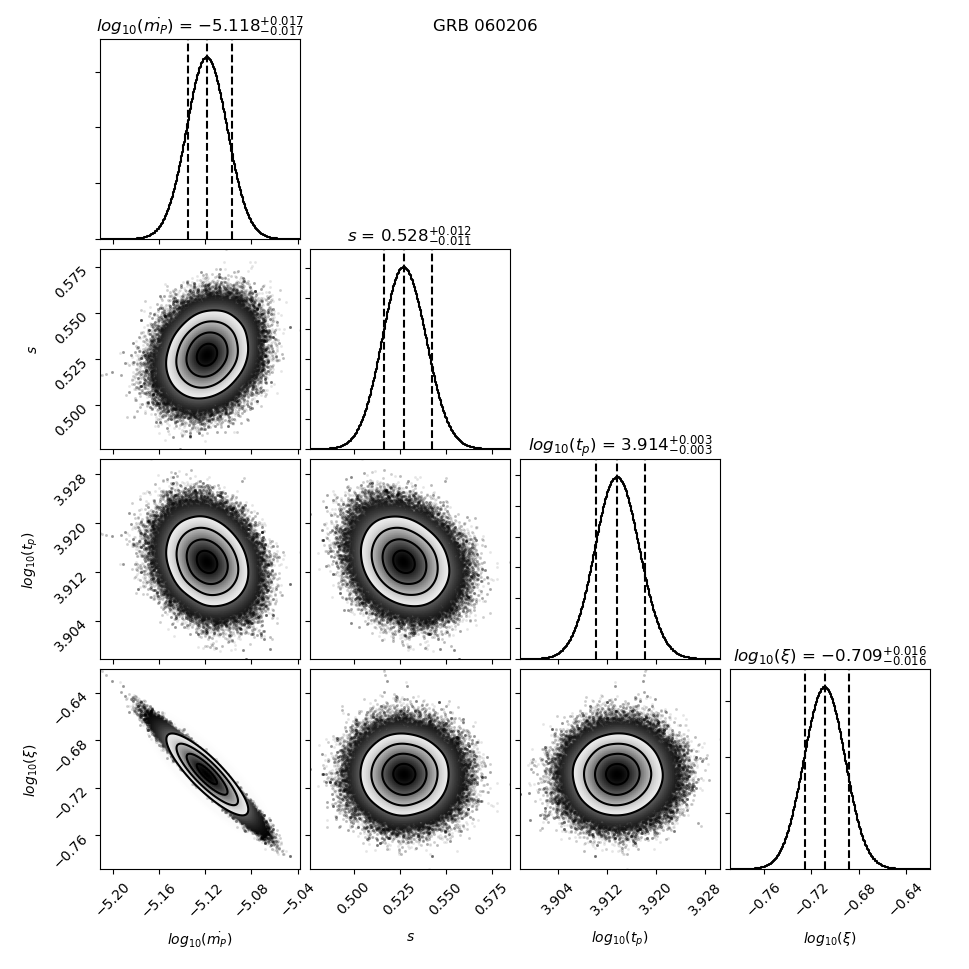}
	   \includegraphics[width=0.32\textwidth,height=6.5cm]{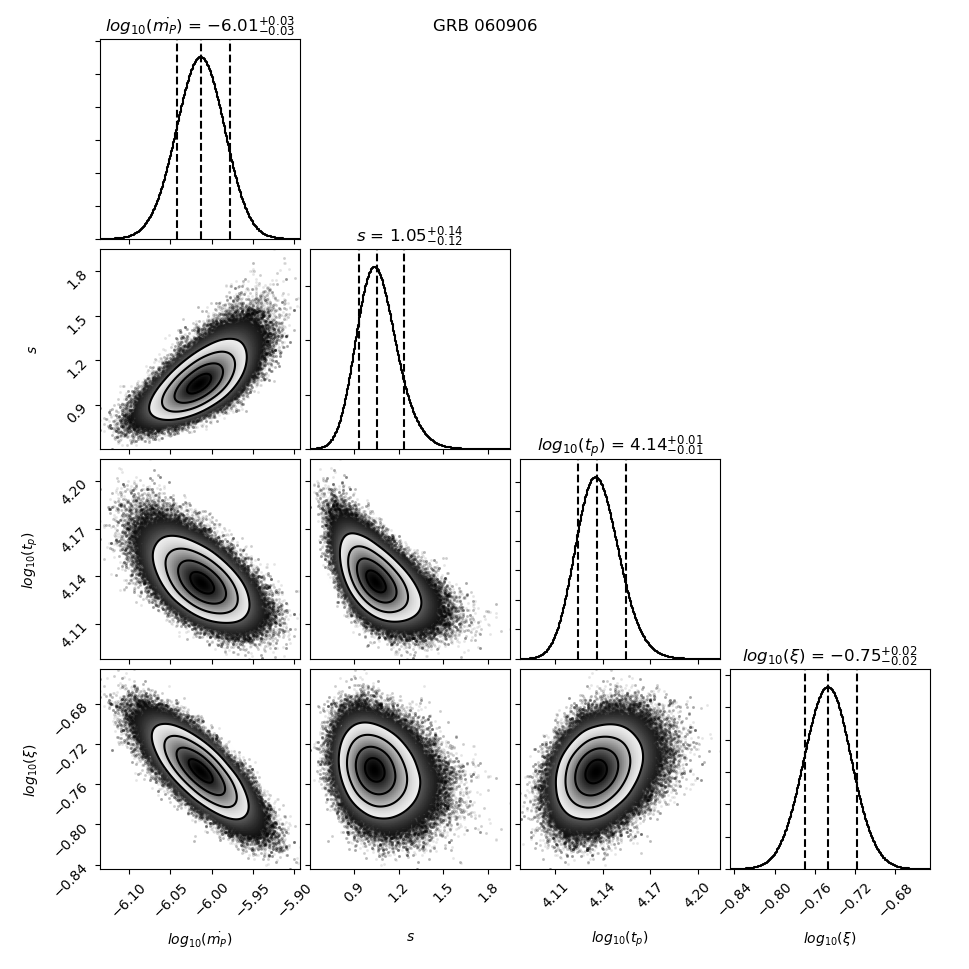}
	   \includegraphics[width=0.32\textwidth,height=6.5cm]{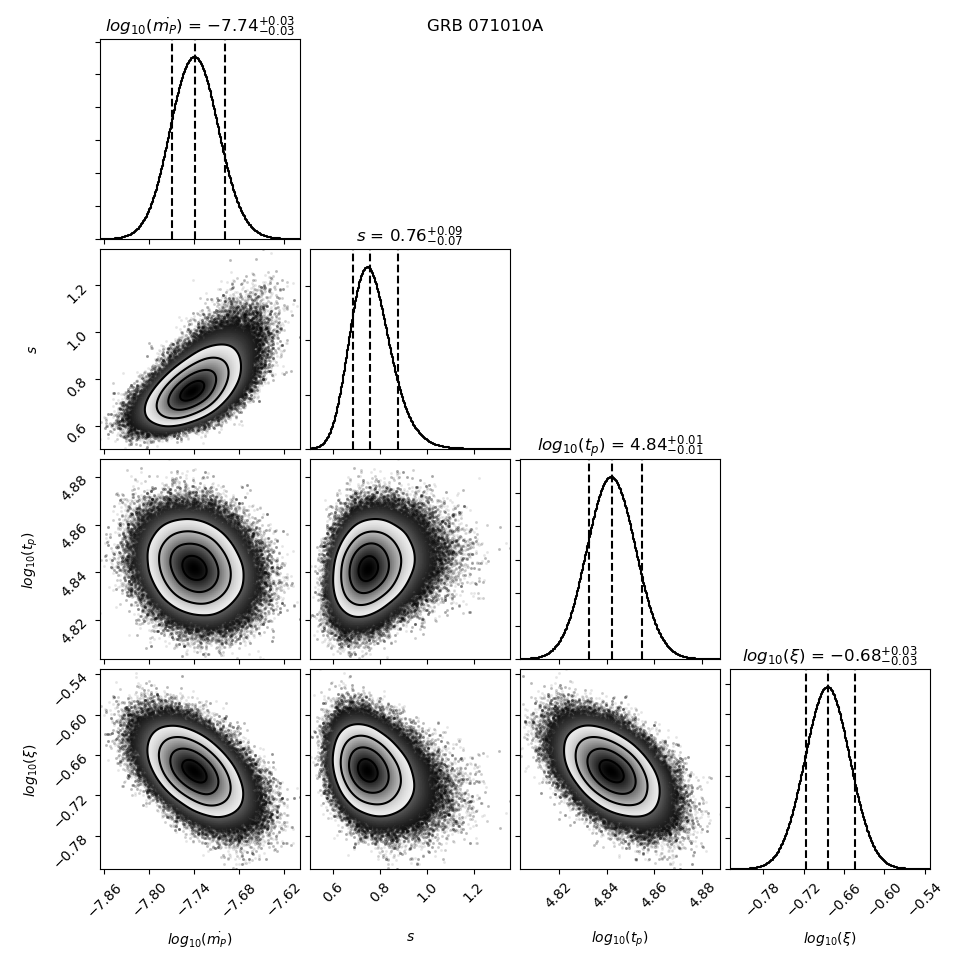}
	   \includegraphics[width=0.32\textwidth,height=6.5cm]{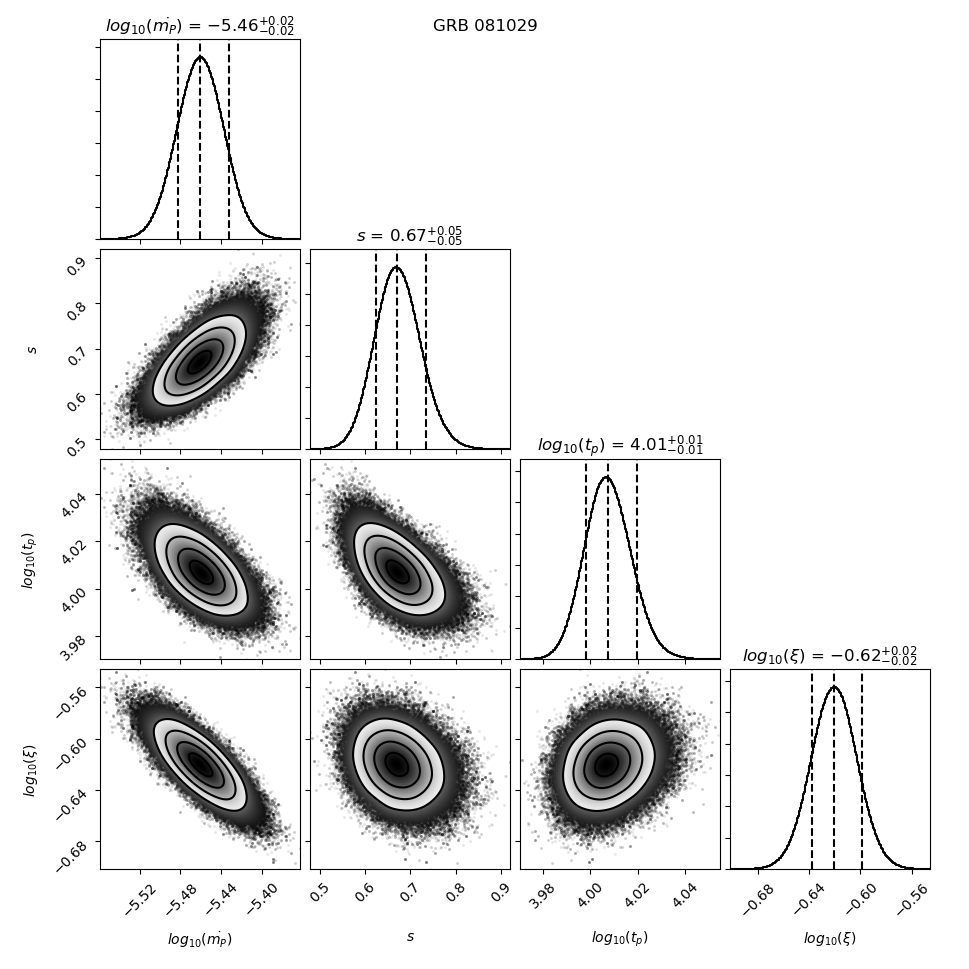}
	   \includegraphics[width=0.32\textwidth,height=6.5cm]{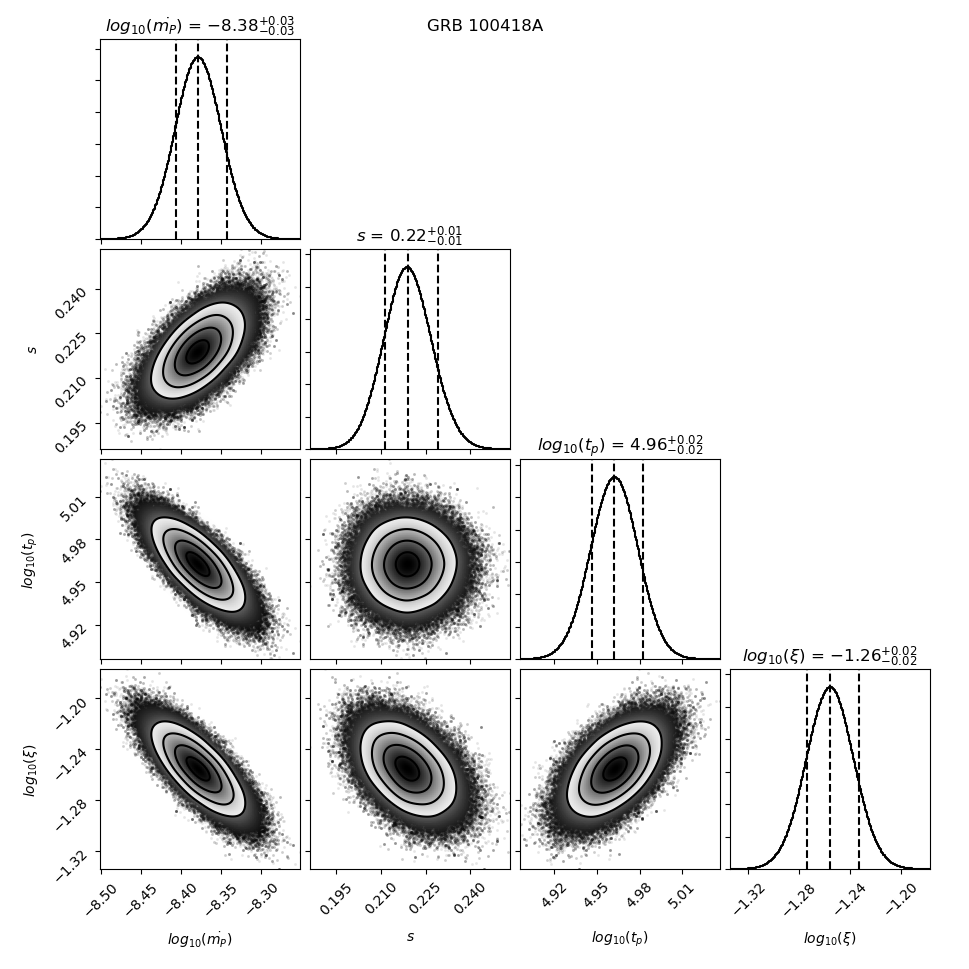}
	   \includegraphics[width=0.32\textwidth,height=6.5cm]{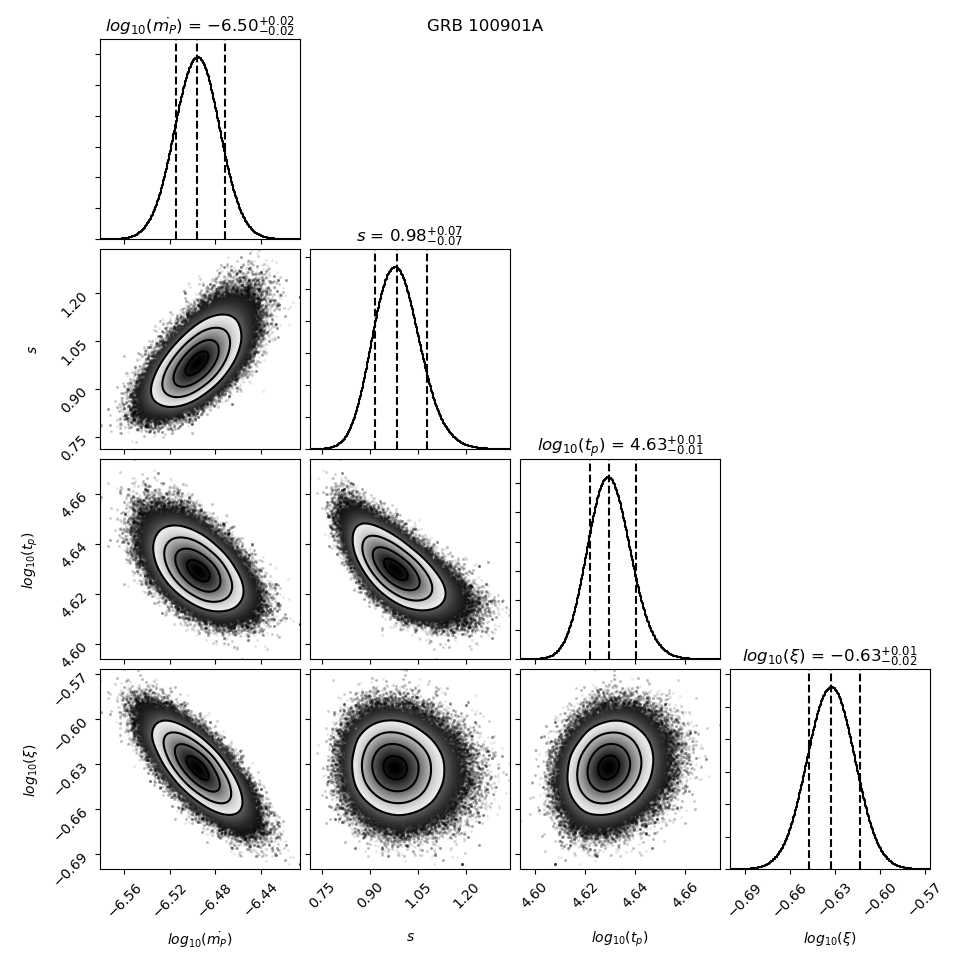}
       \caption{The corner plots  of the free parameters posterior probability distribution for the fitting  of giant X-ray and optical bump in  the gold  sample.} 
	\label{corner-1}
\end{figure}

The best fitted light curves for gold sample are shown in Figure \ref{fittingresult-1} . The corner plots of free parameters posterior probability distribution for the fitting are shown in Figure \ref{corner-1}.  We adopt the median values with 1 $\sigma$ errors level as the fitting results, which are listed in Table \ref{table-2}. We also calculate the total accretion mass $M_{\rm acc}$ during the fall-back process,  the BH magnetic field strength $B_{p,15}$ at  $t_p$ and the fall-back radius $r_p$ corresponding to $t_p$. We find that the X-ray and optical data could be well fitted simultaneously when $\xi$ is in order of $10^{-1}$. It is interesting to note that the optical and X-ray flux ratio is consistent with the standard $F_{\nu}\propto\nu^{1/3}$ synchrotron spectrum below $E_p$. Therefore, we take $\eta_{X}=0.01$ and $\eta_{O}=0.001$ to fit the X-ray or optical bumps in the silver sample.   

The best-fitted light curves for the silver sample are shown in Figure \ref{fittingresult-2}. The corner plots of the free parameter posterior probability distribution for the fitting are shown in Figure \ref{corner-2}. 
MCMC Fitting results for the silver sample are listed in Table \ref{table-3}. It is worth noting that the optical data of GRB 051016B shows a simple power-law decay segment during the X-ray bump epoch. In this case, we can give an upper limit for $\xi<0.015$, which is one order of magnitude lower than the value in the gold sample. For GRB 130831A, there is a simultaneous raising part for both optical and X-ray bands, but unfortunately the X-ray bump signature is incomplete due to the data gap. We test a simultaneous fit for its optical and X-ray data, and find that the $\xi$ value for GRB 130831A would be roughly 0.02, which is also lower than the value in the gold sample. If the giant bumps found in this work are indeed from the internal dissipation of fall-back accretion powered jets, the dissipation ratio between X-ray and optical bands might be diverse. For other bursts in the silver sample, there is no good joint band data to make any interesting constraints on their $\xi$ values.

\begin{figure}[hbt]
       \figurenum{6}
	    \centering
        \includegraphics[width=5.2cm,height=6cm]{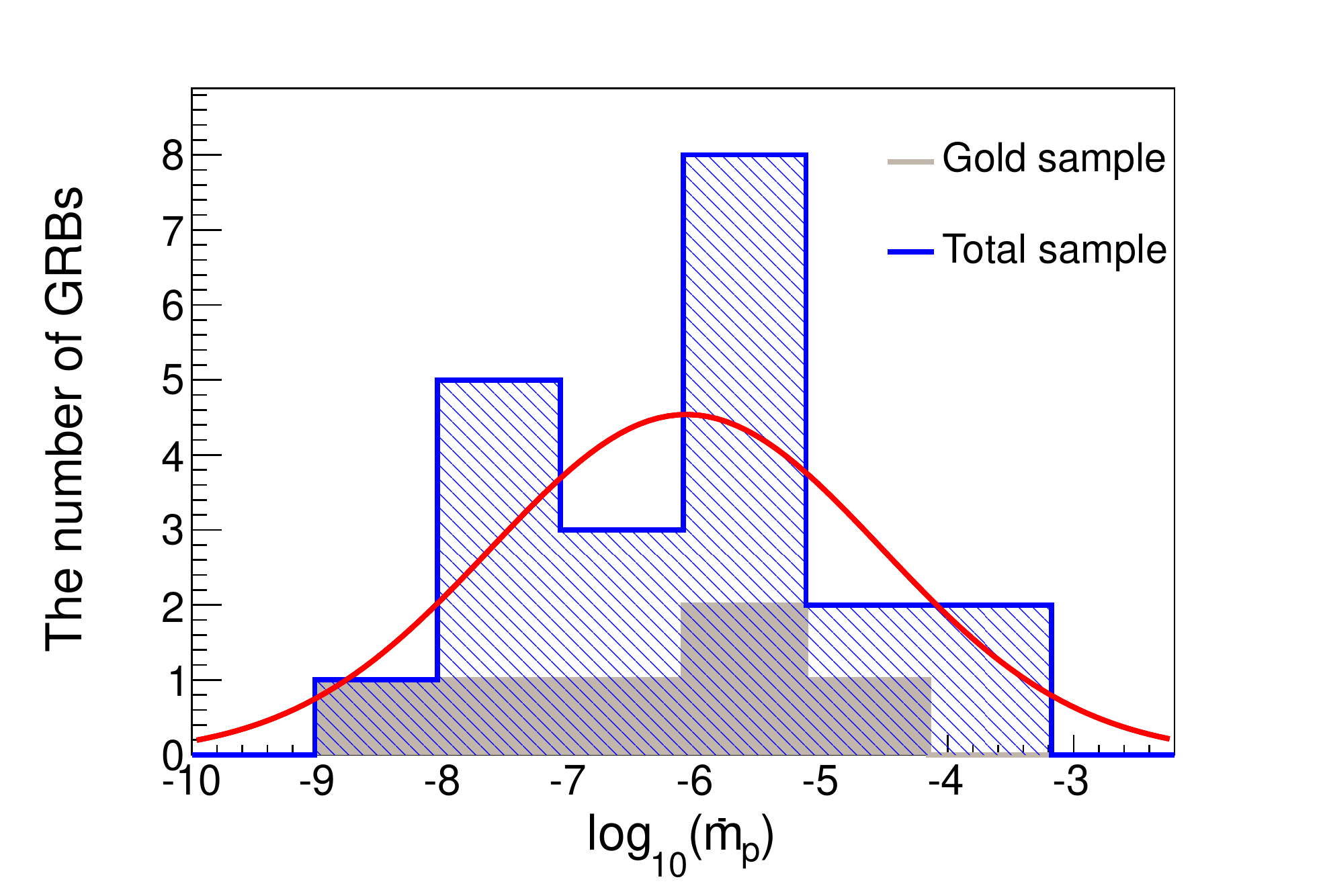}
        \includegraphics[width=5.2cm,height=6cm]{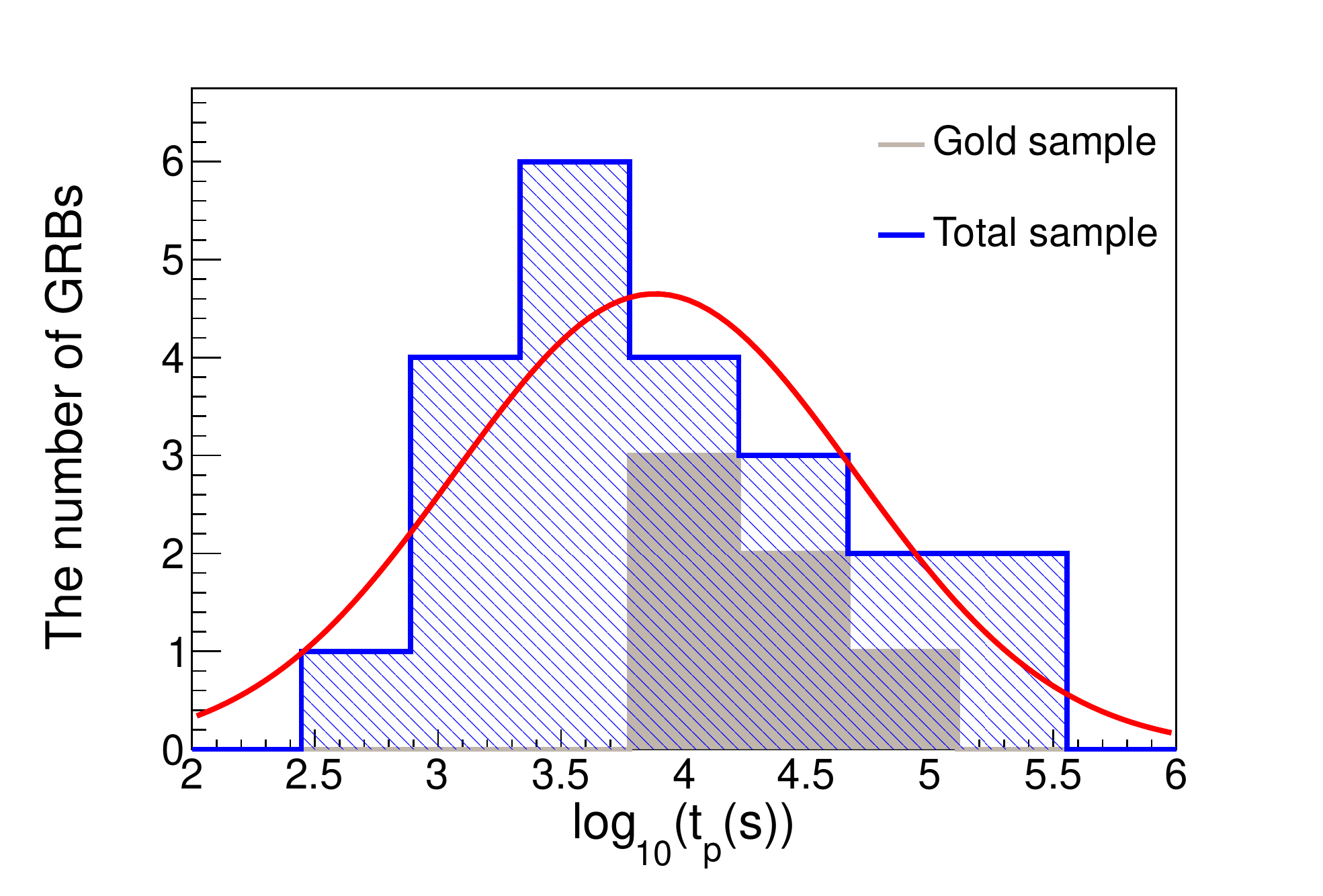}
         \includegraphics[width=5.2cm,height=6cm]{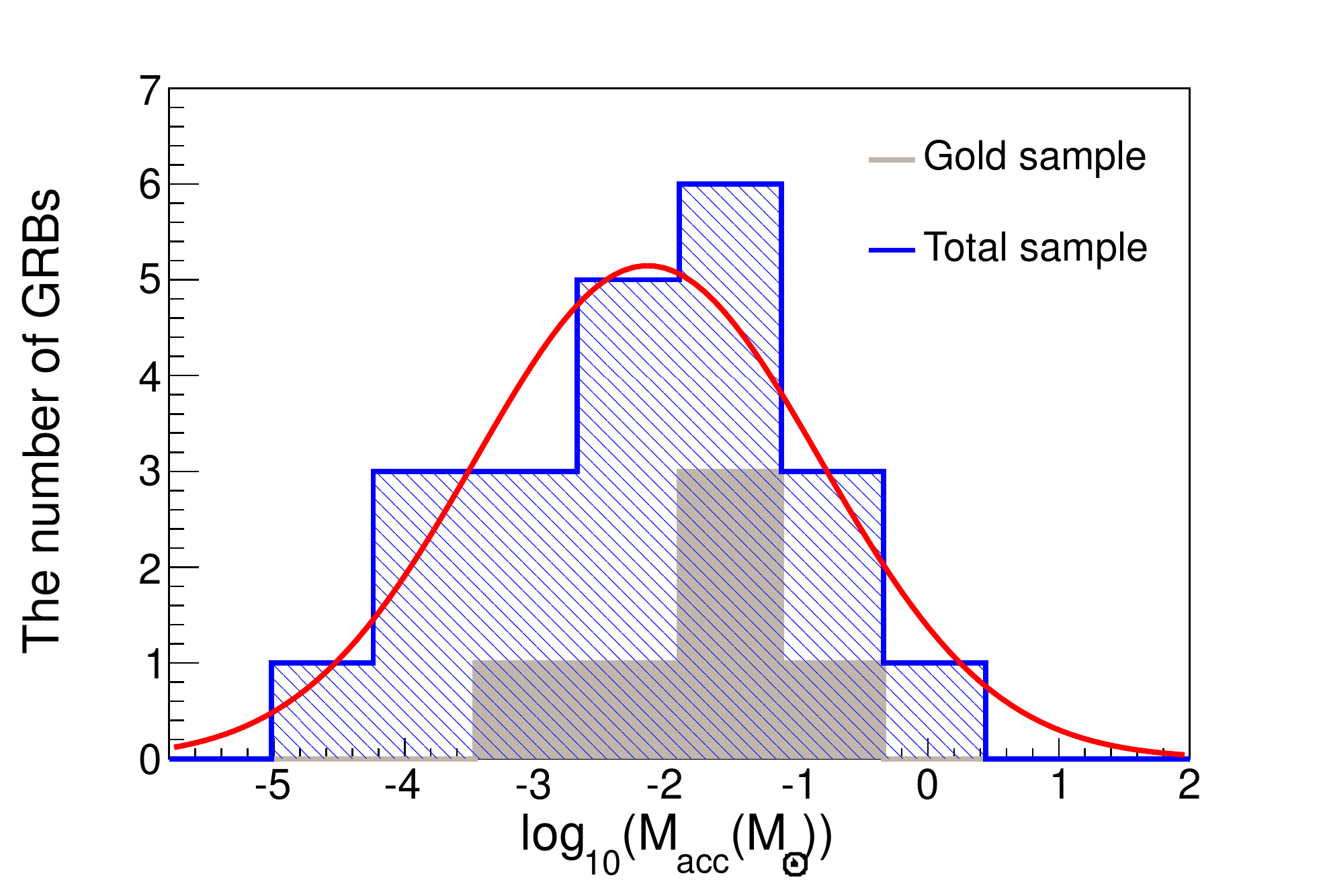}
        \includegraphics[width=5.2cm,height=6cm]{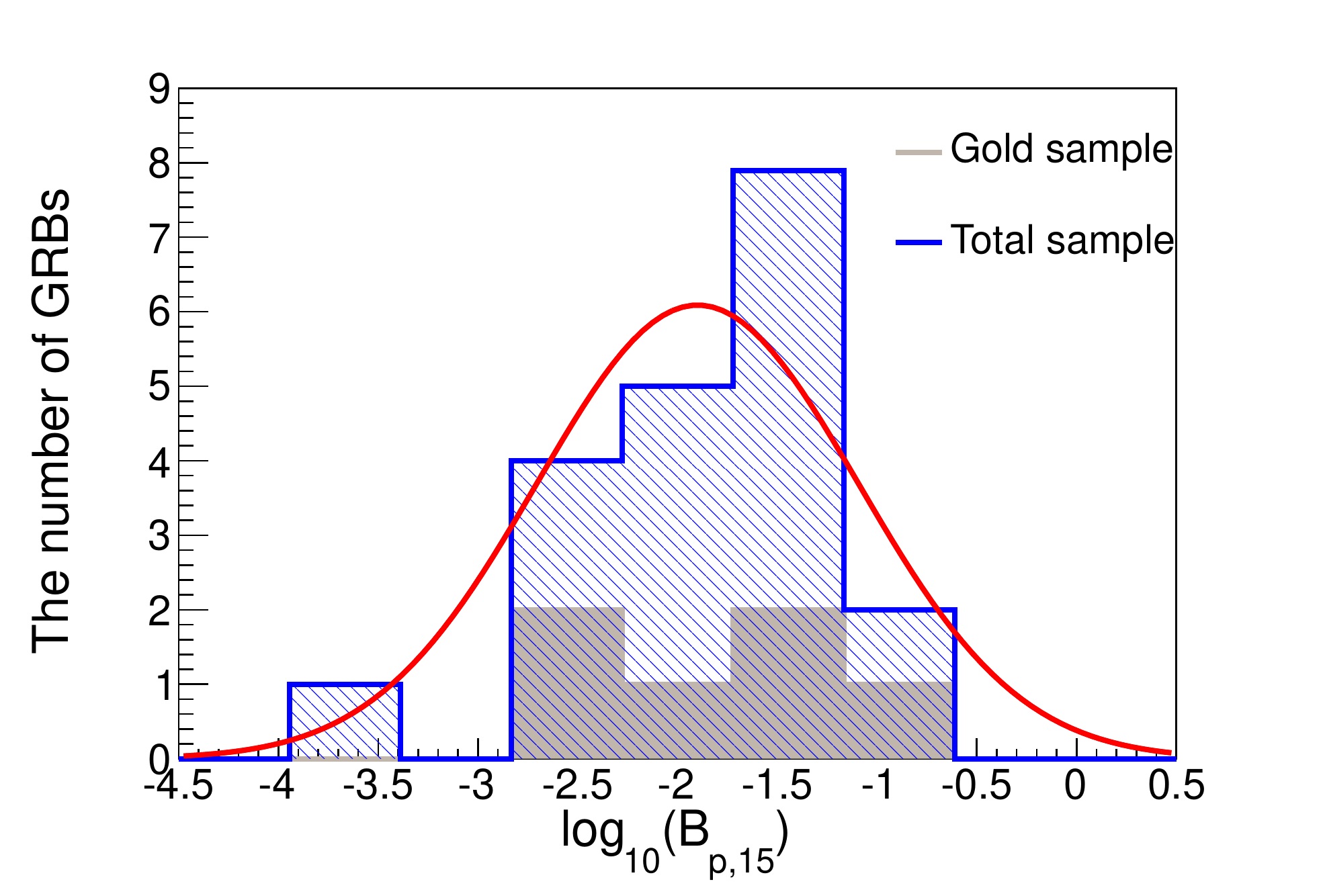}
        \includegraphics[width=5.2cm,height=6cm]{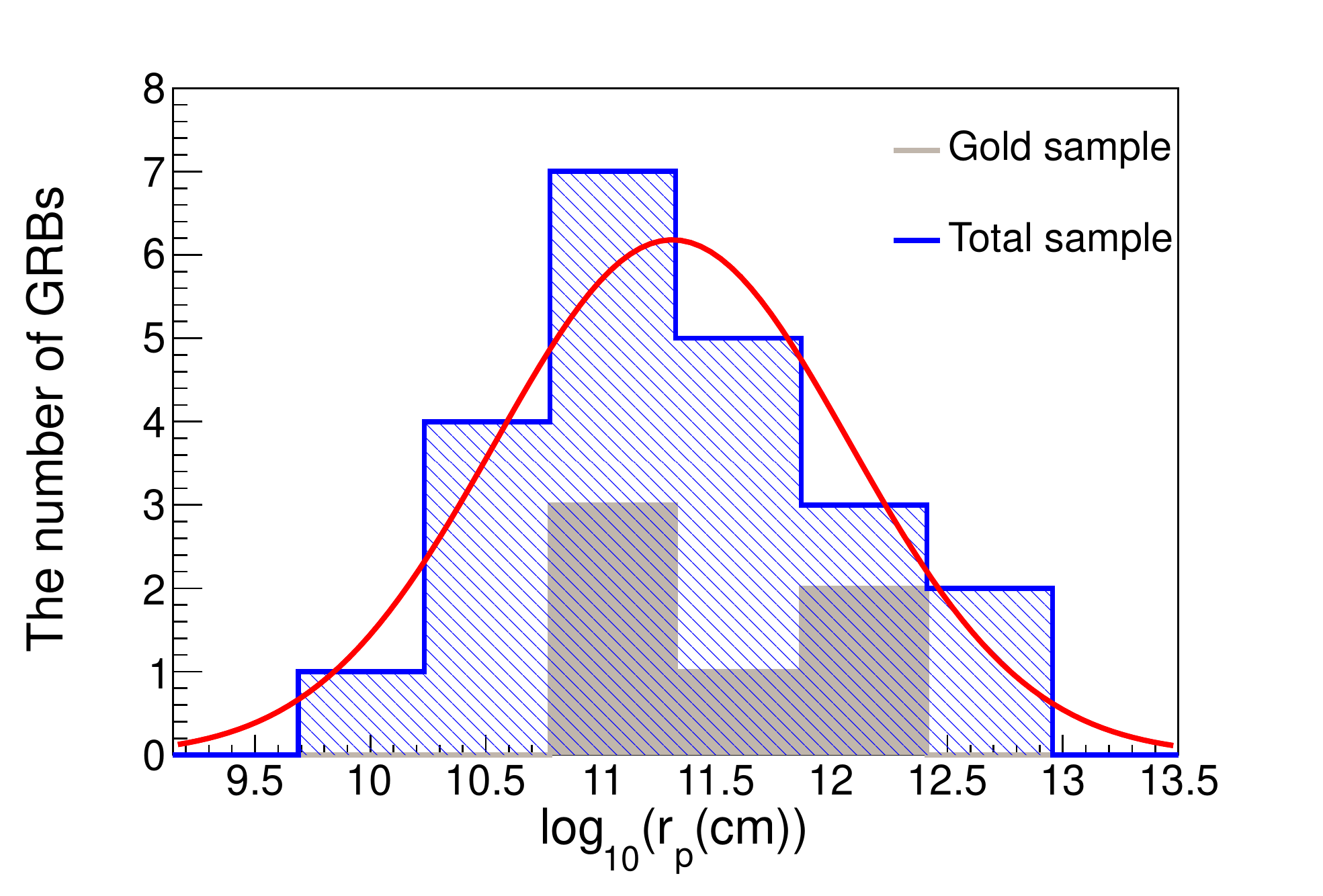}
       \caption{The distributions of $\dot{m}_{p}$, $t_{p}$,  $M_{\rm acc}$  $B_{p,15}$ and $r_{p}$. The  red solid lines are results of Gaussian fits for the total sample.}
	\label{dis}
\end{figure}

We display the distributions of best fitting values of $\dot{m}_{p}$, $t_{p}$,  $M_{\rm acc}$,  $B_{p,15}$ and $r_p$ for both the  gold and total samples in Figure \ref{dis}. We find that for the total sample,  $\dot{m}_{p}$, $t_{p}$,  $M_{\rm acc}$  $B_{p,15}$ and $r_p$ accord the lognormal distribution in the ranges of  $5.13\times10^{-11}<\dot{m}_{p}<6.1\times10^{-4}$, $512.86~\rm s <t_{p}<3.01\times10^{5}~ \rm s$,    $2.22\times10^{-5}~M_{\odot}<M_{\rm acc}<0.9~M_{\odot}$,  
$1.82\times10^{-4}< B_{p,15}<0.23$, and $1.21\times10^{10}~\rm cm<r_{p}<7.66\times10^{12}~\rm cm$,  respectively, with ${\rm log}_{10}(\dot{m}_{p}) =-6.07\pm1.55$, ${\rm log}_{10}[t_{p}(s)]=3.88\pm0.81$, ${\rm log}_{10}[M_{\rm acc}(M_{\odot})]=-2.14\pm1.31 $, ${\rm log}_{10}(B_{p,15})=-1.89\pm0.81$ and ${\rm log}_{10}[r_{p}(\rm cm)]=11.31\pm0.77$. 

We would like to note that during the sample selection, we also find another five GRBs (GRBs 120215A, 120224A,  130807A,150626B and 150911A), which might contain giant X-ray bump signatures but unfortunately without redshift measurements. If we adopt $z = 1$ for those GRBs and also use MCMC method to fit the giant X-ray bumps for those GRBs, we find that the distributions for all the parameters barely change by adding those GRBs into the sample.

Based on the fitting results, we can draw conclusions as follows: 1) a reasonable  parameter space can interpret  the giant  X-ray and optical bumps for both the gold and silver GRB samples. 2) the lower limit of total accretion mass could be as low as $\sim 10^{-5} M_{\odot}$, which means even with a very small fraction of the progenitor's envelop falling back, it is still possible to generate the giant X-ray and optical bump feature. 3) when the fall-back mass rate reaches its peak, the corresponding radius $r_p$ is around $10^{10}-10^{12}$ cm, which is consistent with the typical radius of a Wolf-Rayet star.

\begin{figure}[hbt]
         \figurenum{7}
	\centering
	\includegraphics[width=12cm,height=8cm]{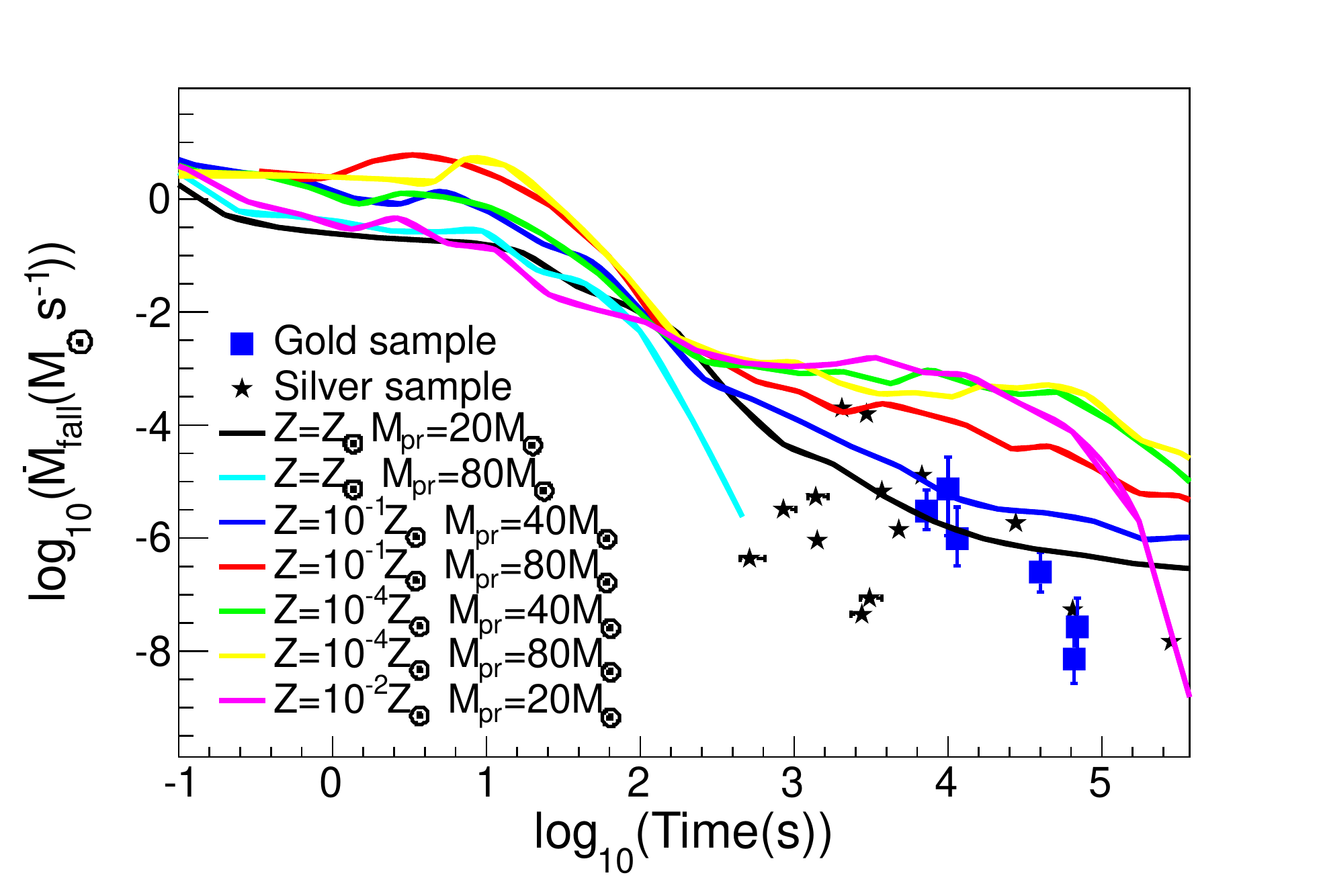}
	\caption{The evolution of the mass supply rate of progenitor with different masses and metallicities. Blue square and black star symbols  represent the best fitting required dimensionless peak accretion rate $\dot{m}_{p}$ with respect to the peak accretion time $t_{p}$. The  $Z_{\odot}$ is the  metallicities of the sun. }
	\label{mdot}
\end{figure}

It is worth checking whether the mass supply rate of progenitor envelope material at $t_{p}$ could meet the accretion rate requirement for our fitting. We calculate the mass supply rate of the progenitor envelope materials as \citep{suwaioka11,woosley12,matsumoto15,liu18}
\begin{eqnarray}
\dot{M}_{\rm fall}=\frac{dM_{r}}{dt_{\rm fb}}=\frac{dM_{r}/dr}{dt_{\rm fb}/dr}=\frac{2M_{r}}{t_{\rm fb(r)}}\left(\frac{\rho_r}{\bar{\rho}-\rho_r}\right),
\label{eq-1}
\end{eqnarray}
where $\rho_r$ is the mass density at radius $r$, $\bar{\rho}=3M_{r}/4\pi r^{3}$ is the average mass density within $r$, $t_{\rm fb(r)}\sim (\pi^{2} r^{3}/8GM_{r})^{1/2}$ is freefall timescale at $r$, and
\begin{eqnarray}
M_{r}=M_{0}+\int_{r_{0}}^{r} 4 \pi r^{2} \rho d r,
\label{eq-1a}
\end{eqnarray}
is the total mass within $r$. We assume that the jet is launched when the  central accumulated mass reach the initial mass of BH ($M_{0}$) and  set $r=r_{0}$ and $t=0$ at this time, i.e. $t=t_{\rm fb}(r)-t_{\rm fb}(r_{0})$. On the other hand, \cite{liu18} gave the relationship between the density $\rho$  and radius $r$ for progenitor with  different masses and metallicities. We thus calculate the evolution of the mass supply rate for various progenitor masses and metallicities. As shown in Figure \ref{mdot}, we find that for most cases provided in \cite{liu18}, the  mass supply rate of progenitor envelope material at $t_{p}$ is large enough to meet the accretion rate requirement for our fitting. In other words, the constraints set by the progenitor's mass and metallicity are not strong, so that the interpretation presented in this work should be justified.

\begin{figure}[hbt]
         \figurenum{8}
	\centering
	\includegraphics[width=12cm,height=8cm]{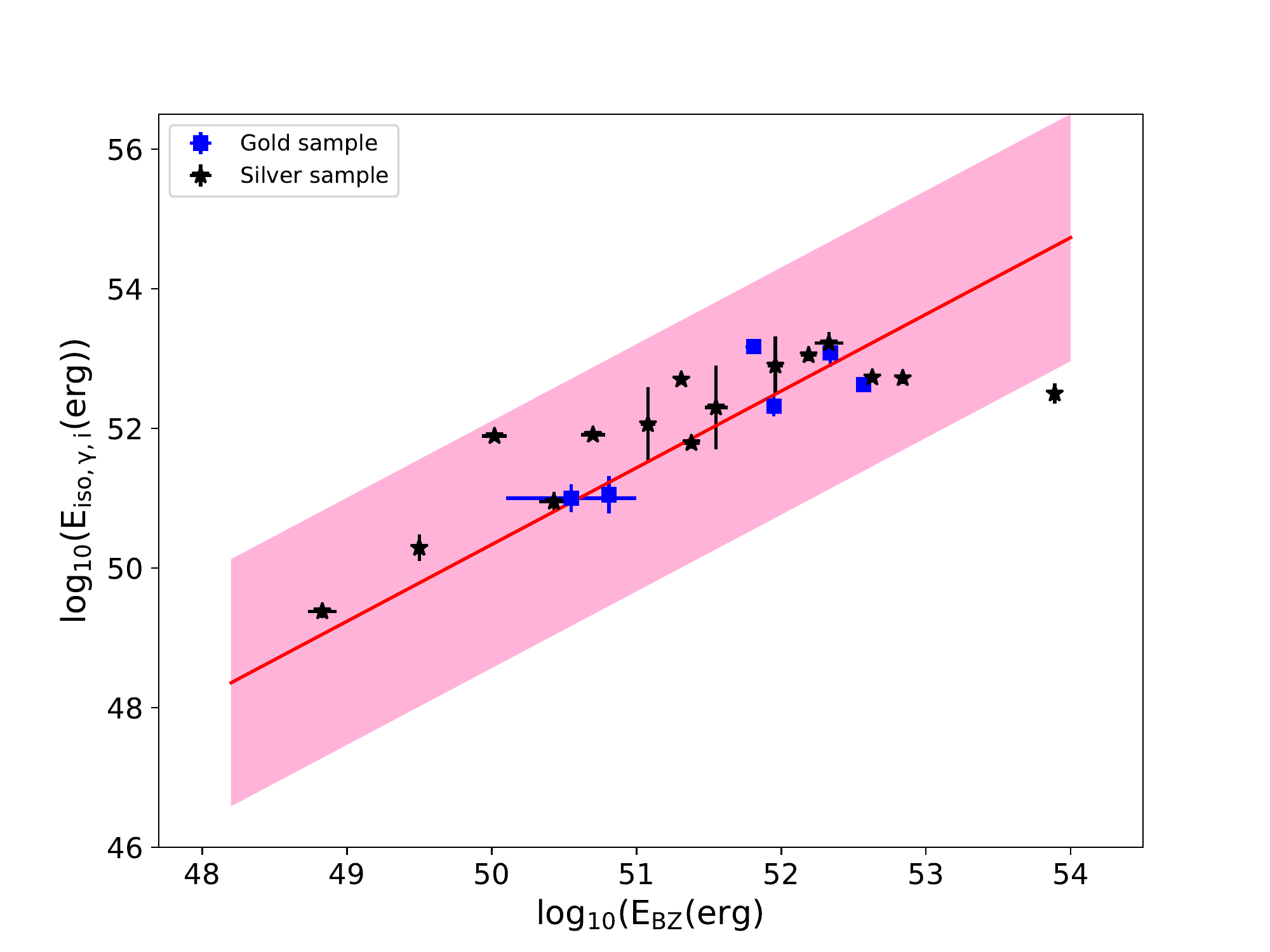}
	\caption{ The relationship between the jet energy from the fall-back accretion ($E_{\rm BZ}$) and isotropic-equivalent radiation energies of GRB prompt phase in the $1-10^{4}$ keV band ($E_{\rm iso,\gamma,i}$). The red line marks the best fitting result and the pink shadowed region shows the intrinsic scatter to the population 3$\sigma$. }
	\label{Eisogamma}
\end{figure}

It is also of interest to study the relation between the jet energy from the fall-back accretion ($E_{\rm BZ}$) and isotropic-equivalent radiation energies of GRB prompt phase in the $1-10^{4}$ keV band ($E_{\rm iso,\gamma,i}$). As shown in Figure \ref{Eisogamma}, we apply the regression model to our sample and obtain the best fit as $\log_{10}(E_{\rm iso,\gamma,i})=(14.41\pm0.57)+(0.73\pm0.11)\log_{10}(E_{\rm BZ})$ (Spearman correlation coefficient $r = 0.82$ and significance level $p<10^{-4}$ for $N = 21$). The result infers that the fall-back accretion is correlated with the prompt phase accretion. 

Finally, we would like to note that besides the internal dissipation, a good fraction of the fall-back jet energy would eventually inject into the GRB afterglow blast wave. Depending on the ratio between the injected energy and the initial kinetic energy in the blast wave, the afterglow light curve 
following the giant bump could become shallower if the late injected energy is larger, or not shallower if the initial kinetic energy is larger \citep{zhao20}. Late time jet break effect could make the situation even more complicated. In our sample, for GRBs 050814, 051016B, 081028 and 140515A, we find that the segment following the X-ray bump tends to become shallower, but not for other bursts.

\begin{figure}[hbt]
        \figurenum{4}
	    \centering
         \includegraphics[width=5.86cm,height=6.5cm]{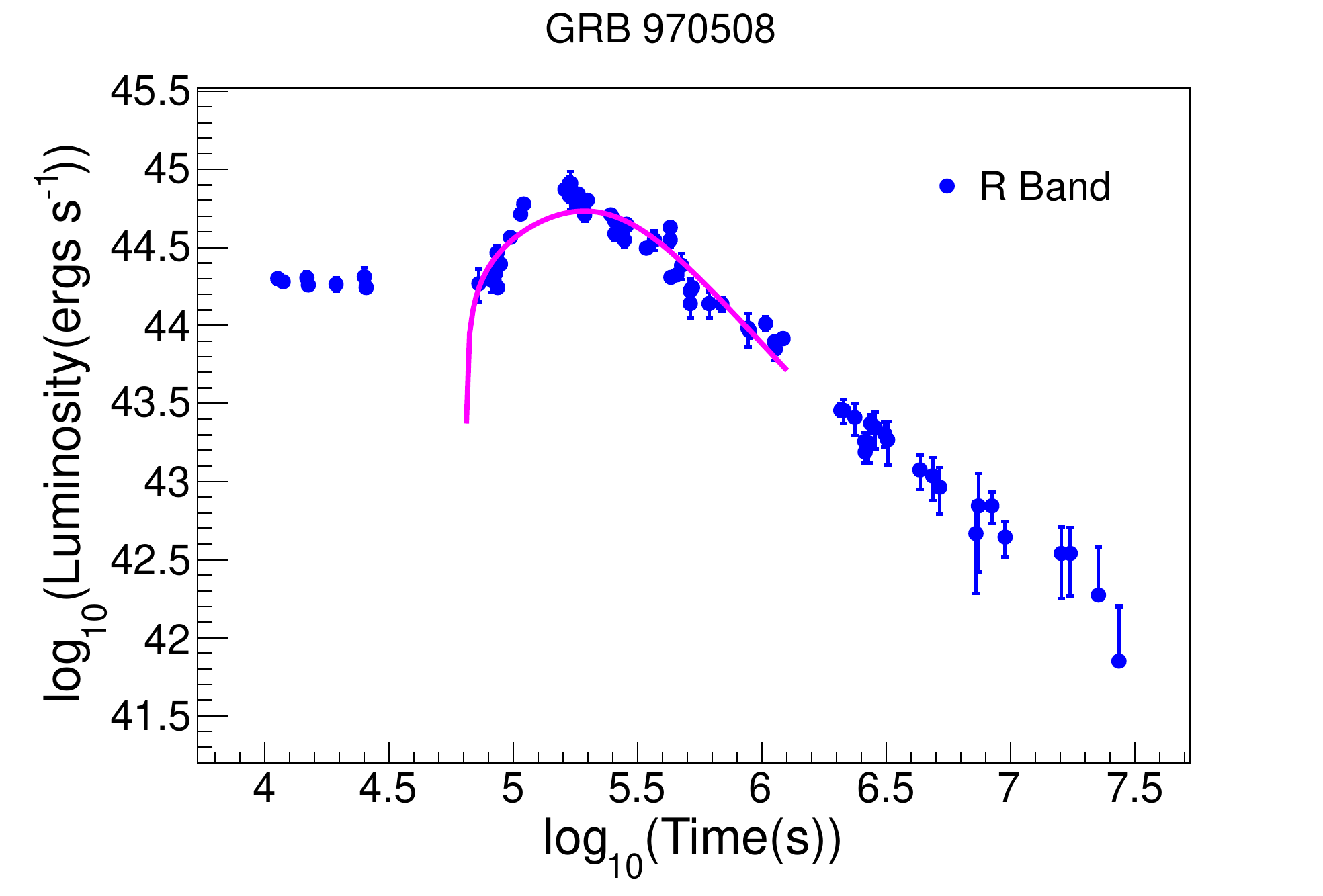}
         \includegraphics[width=5.86cm,height=6.5cm]{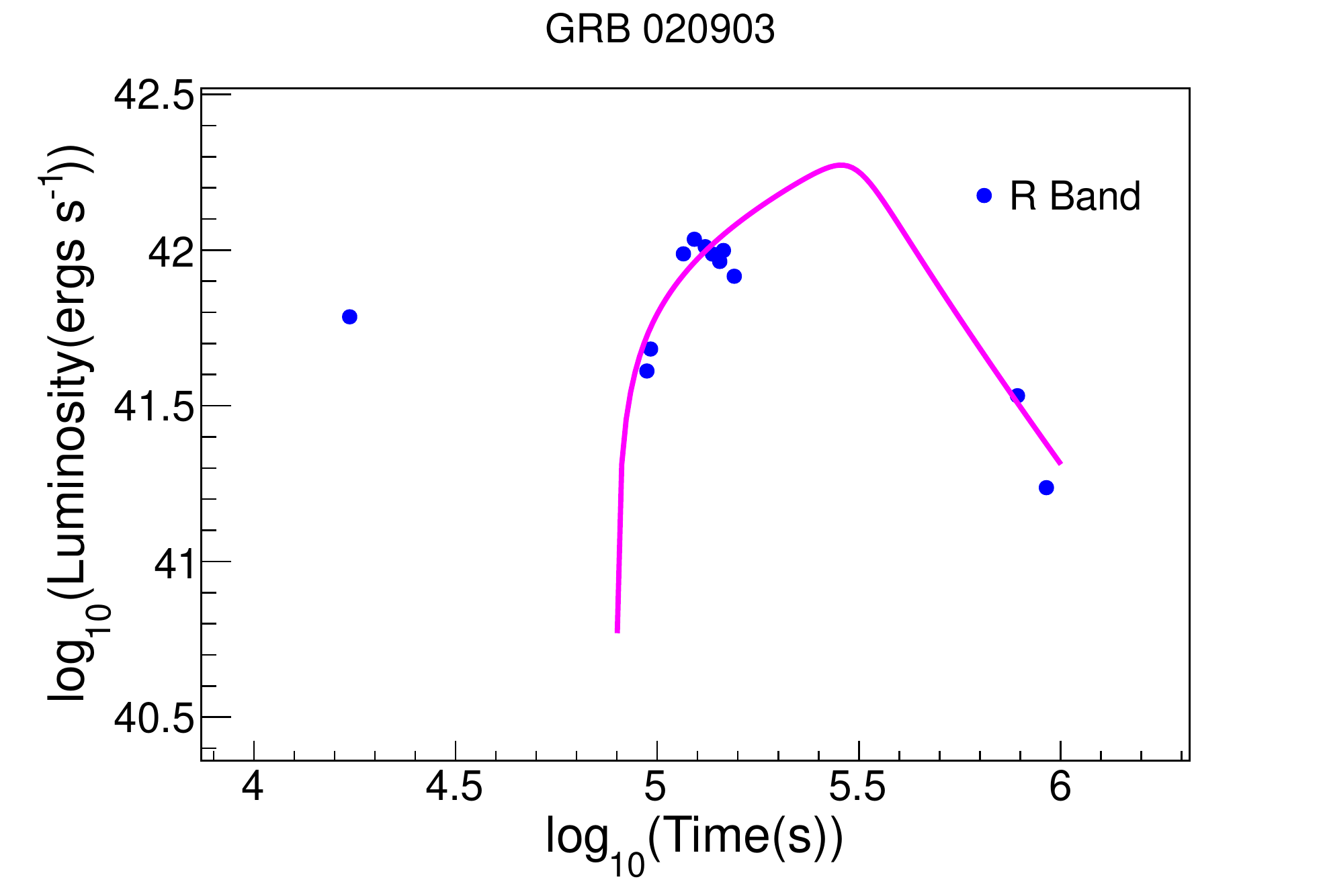}
         \includegraphics[width=5.86cm,height=6.5cm]{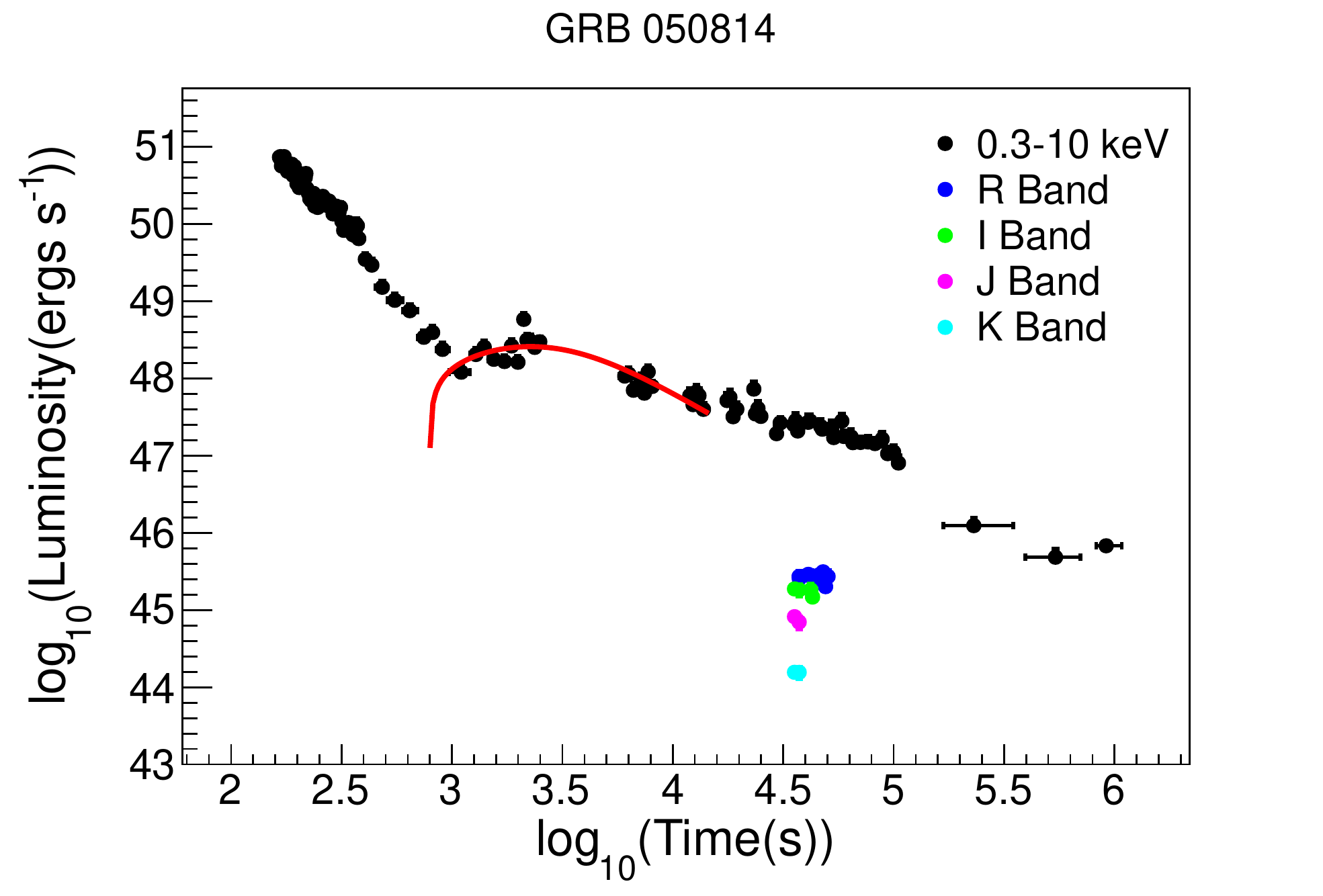}
         \includegraphics[width=5.86cm,height=6.5cm]{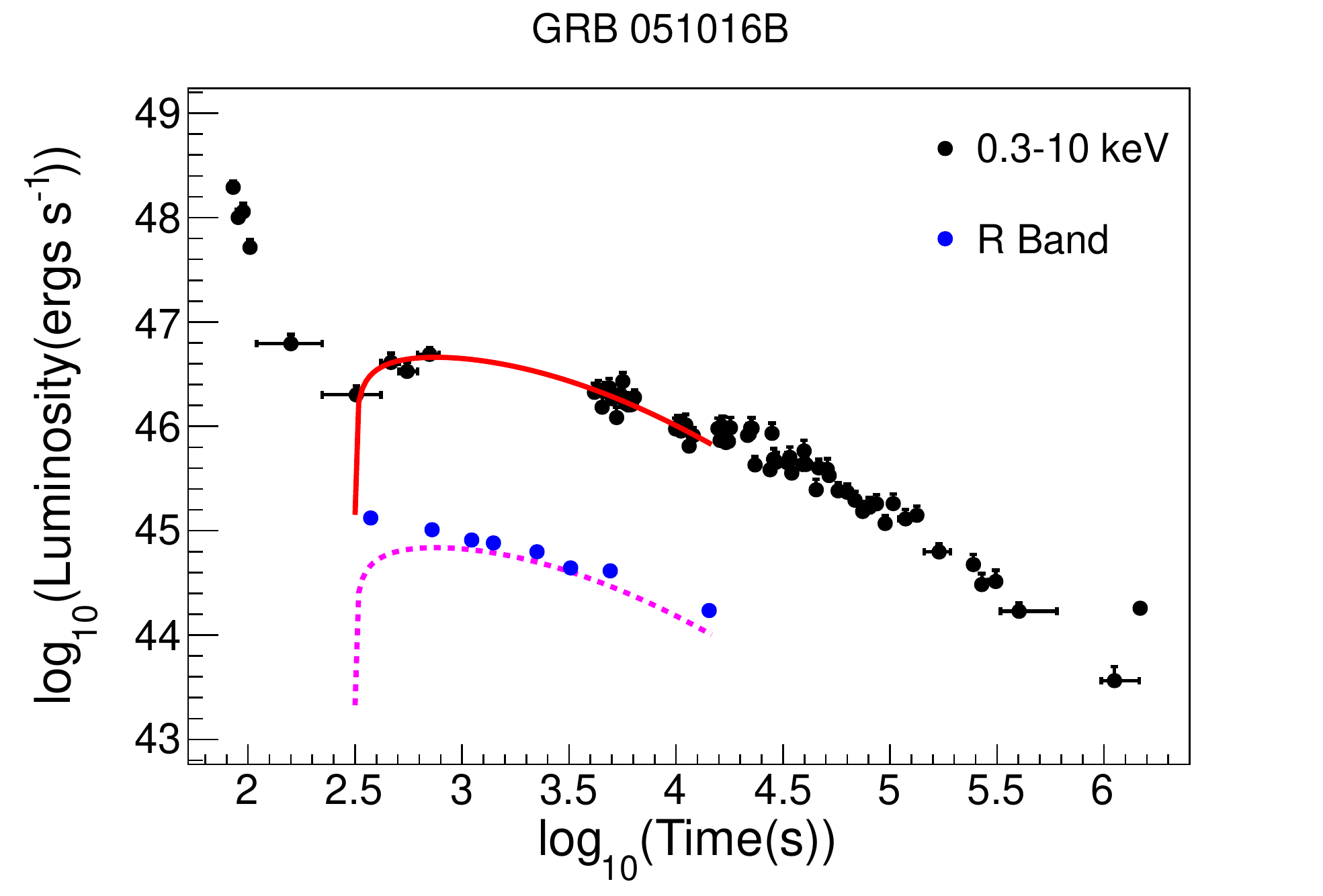}
         \includegraphics[width=5.86cm,height=6.5cm]{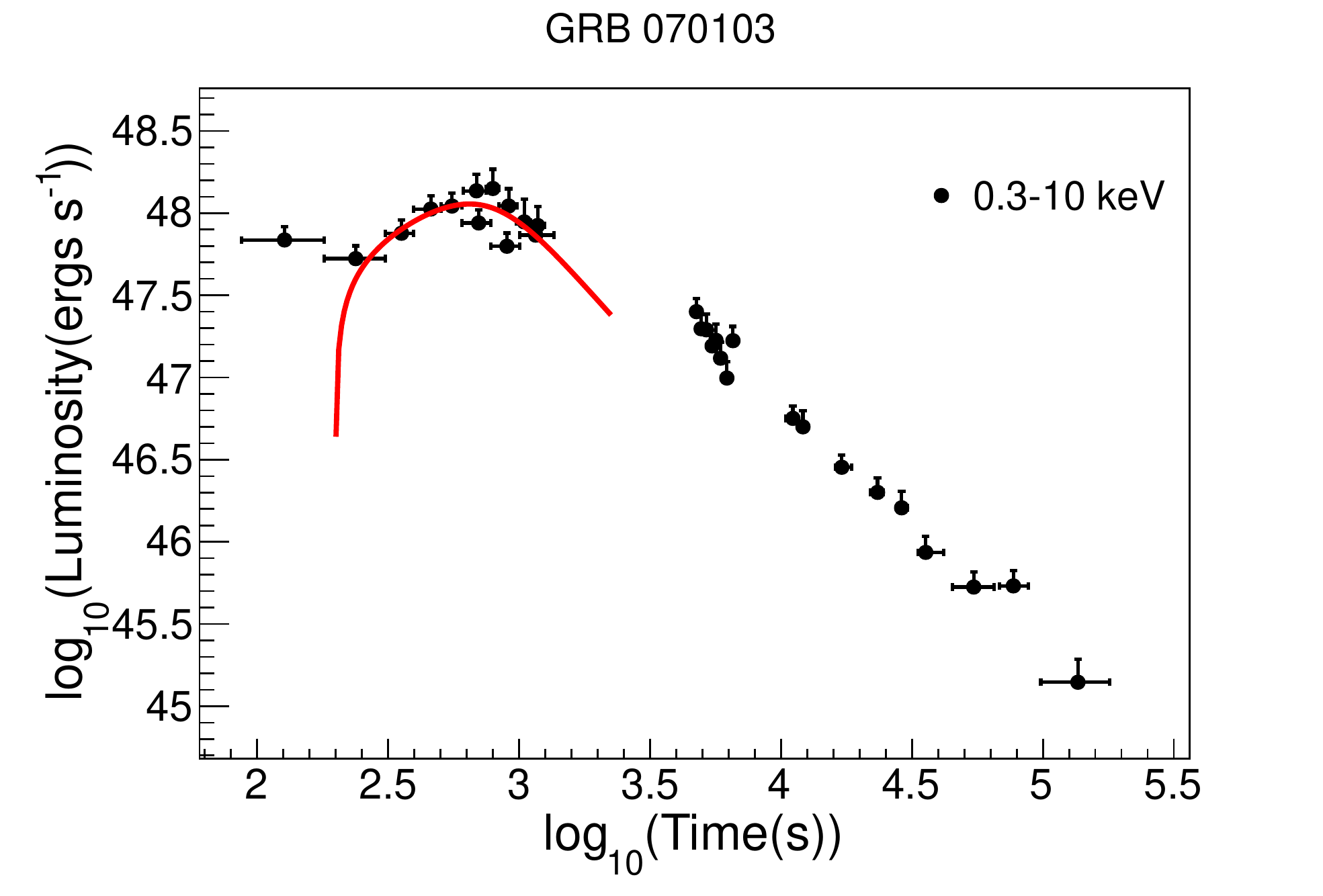}
         \includegraphics[width=5.86cm,height=6.5cm]{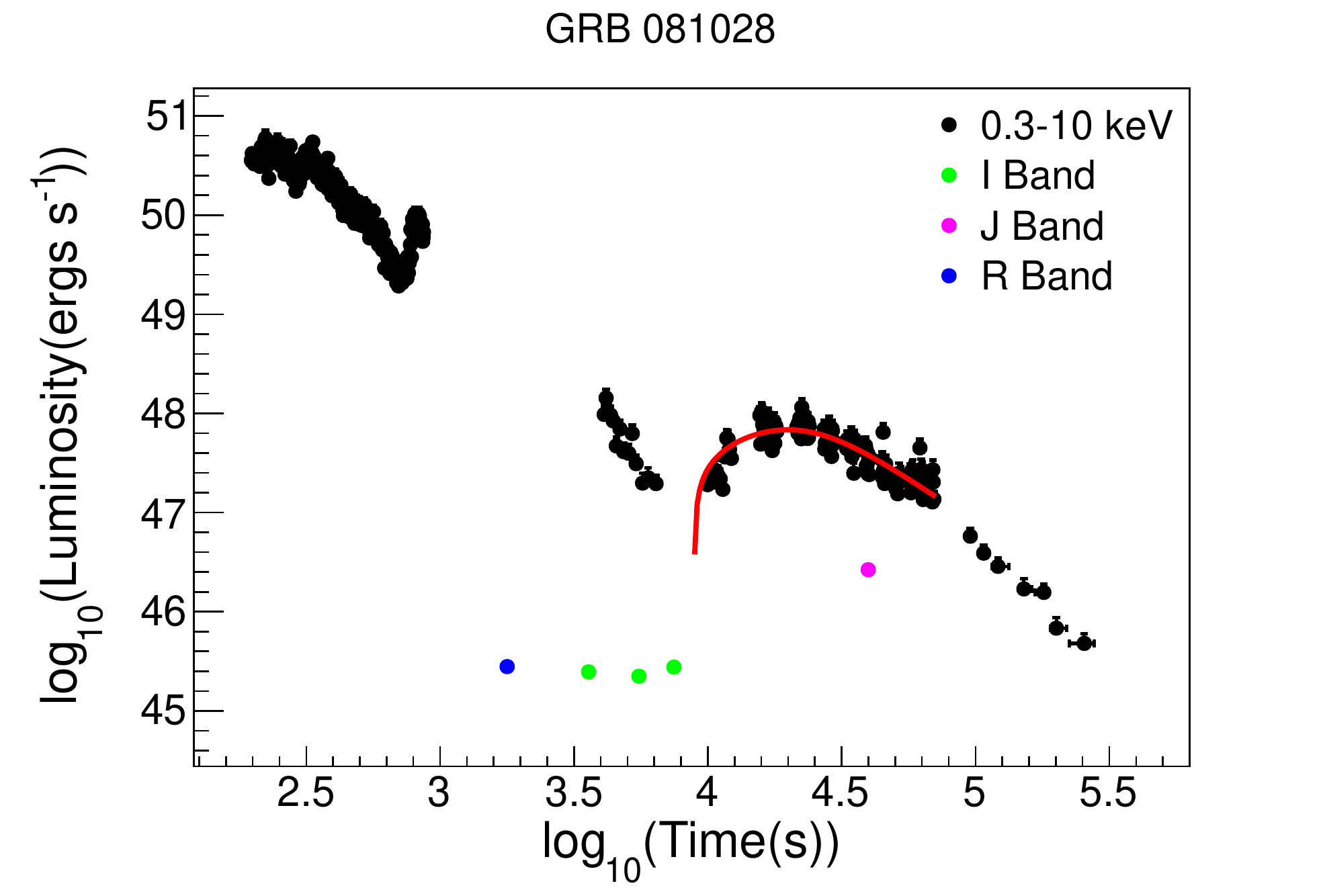}
         \includegraphics[width=5.86cm,height=6.5cm]{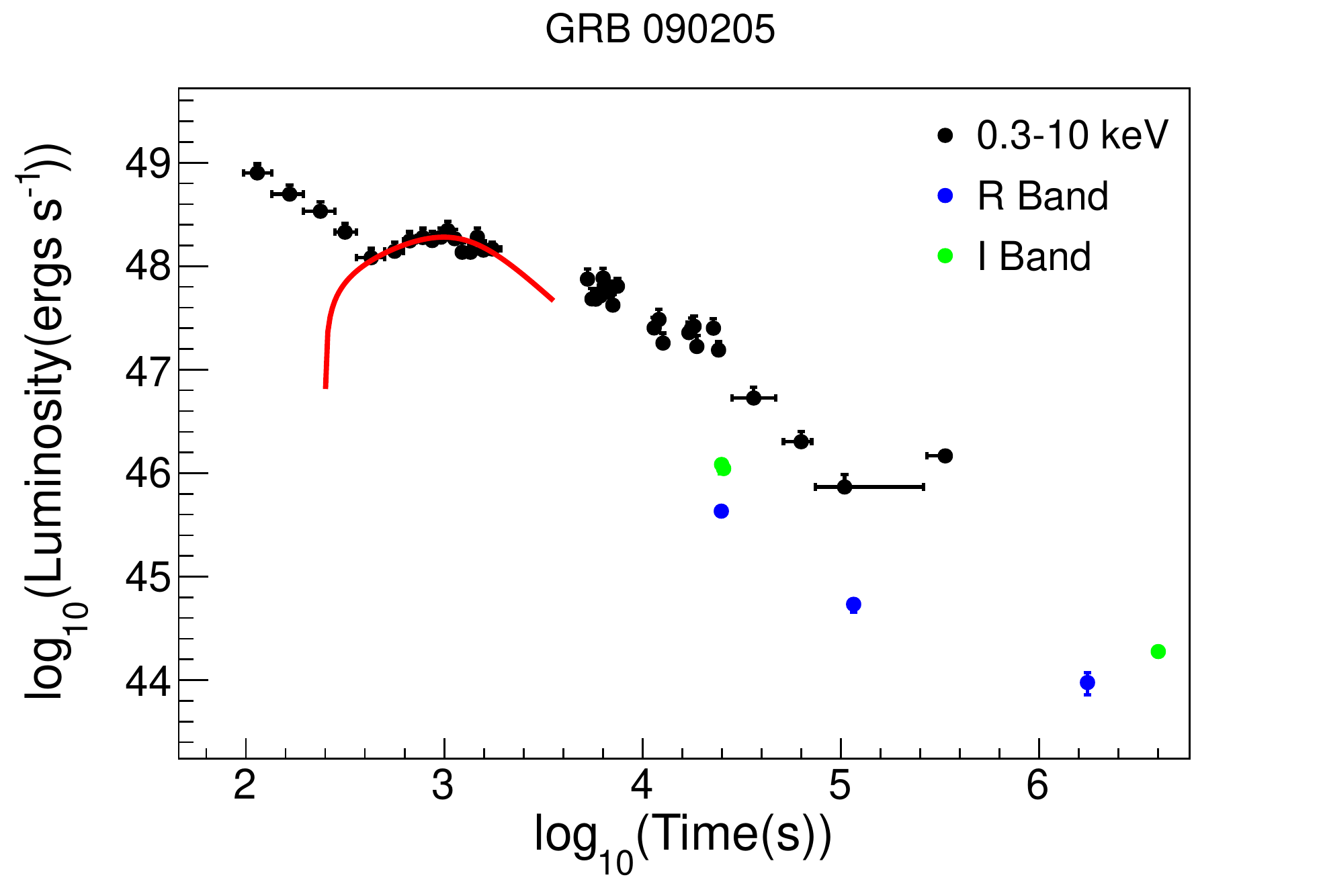}
         \includegraphics[width=5.86cm,height=6.5cm]{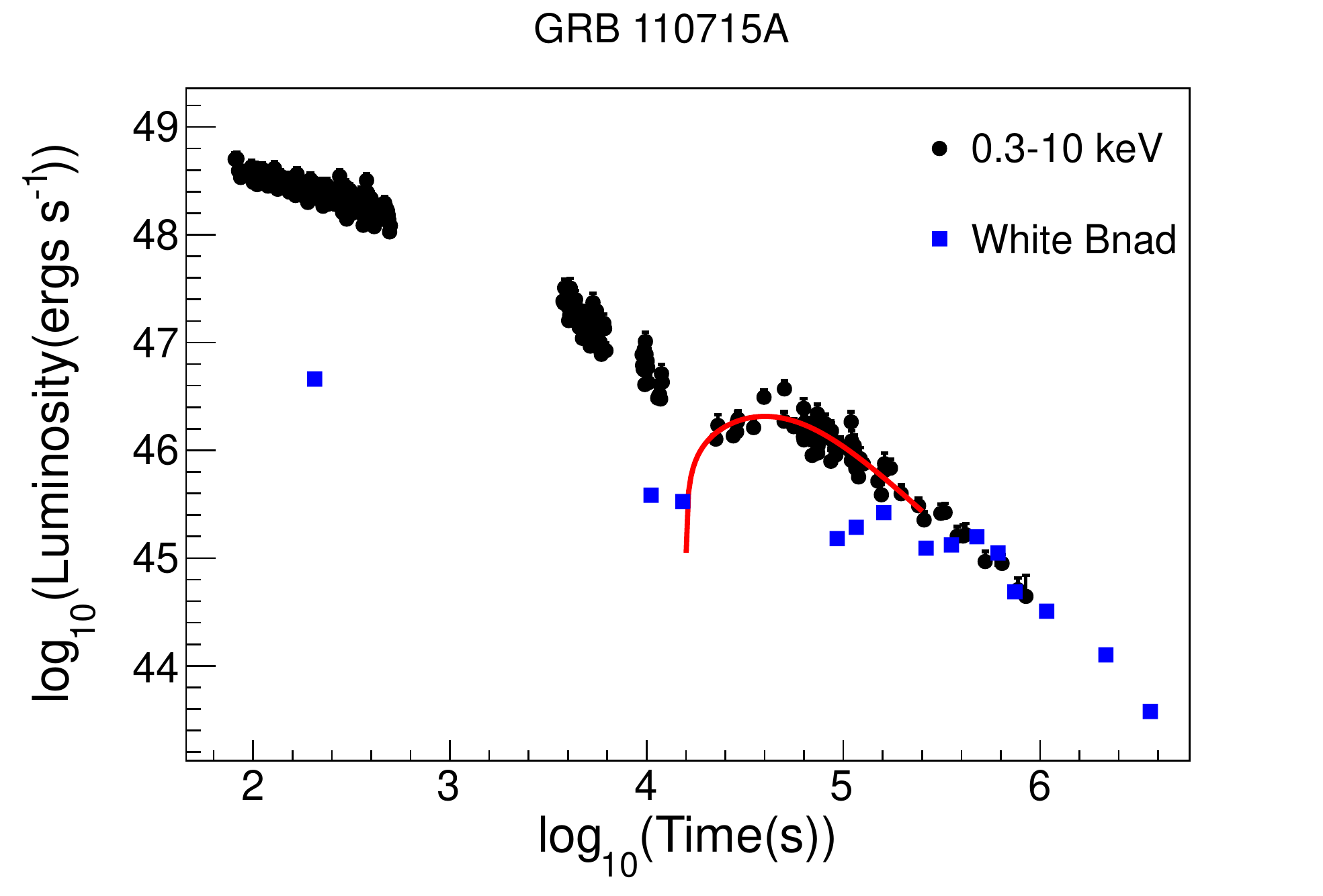}
         \includegraphics[width=5.86cm,height=6.5cm]{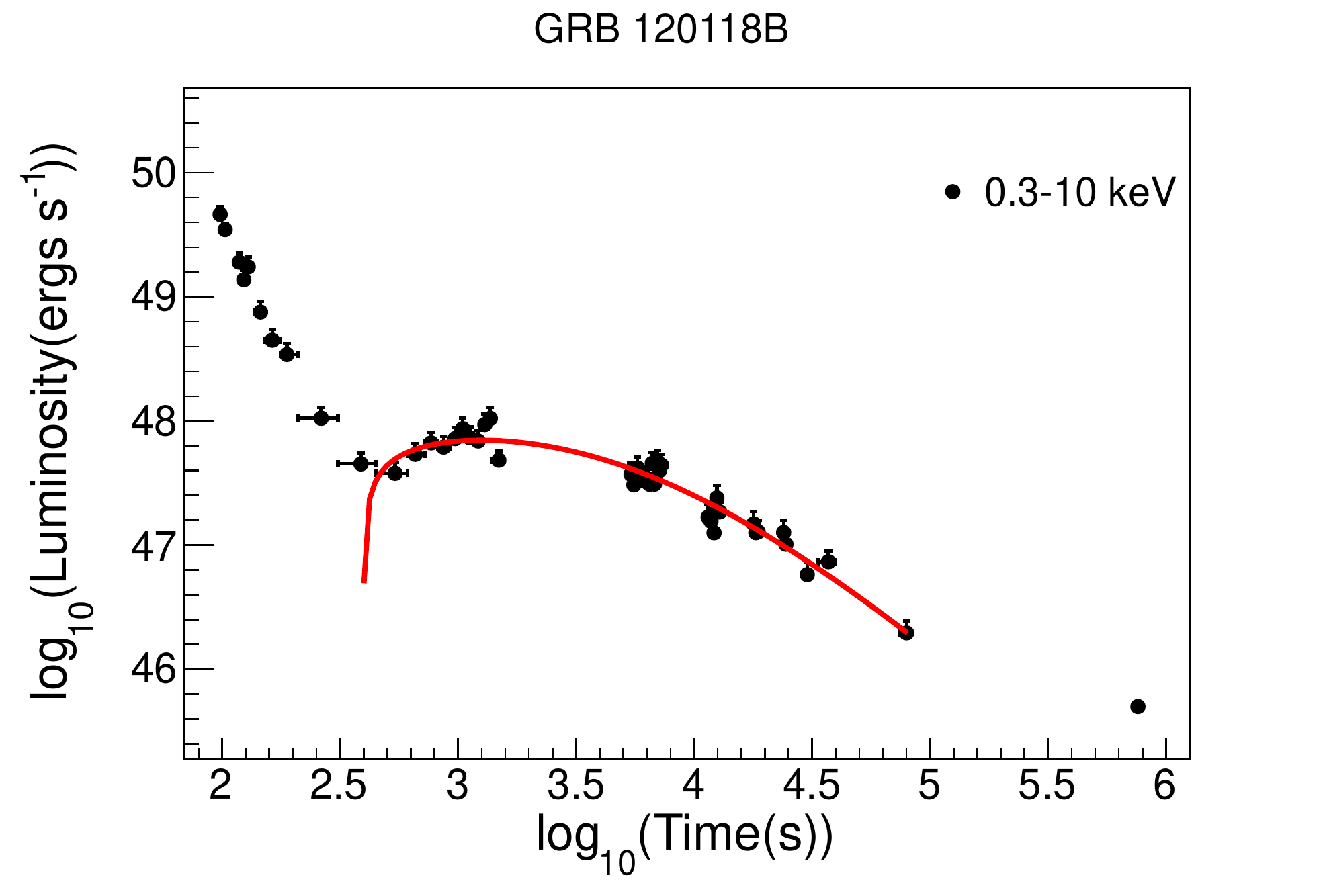}
         \caption{Modeling results for  the giant X-ray or optical bumps of the silver sample. The red solid  lines and pink  solid  lines  present a  theoretical light curve  produced by our model within the the Swift-XRT energy band ($0.3-10$ keV) and optical band, respectively. The pink dotted lines present the upper limit $\xi$. }
	     \label{fittingresult-2}
\end{figure}

\begin{figure}[hbt]
        \figurenum{4}
	    \centering
        \includegraphics[width=5.86cm,height=6.5cm]{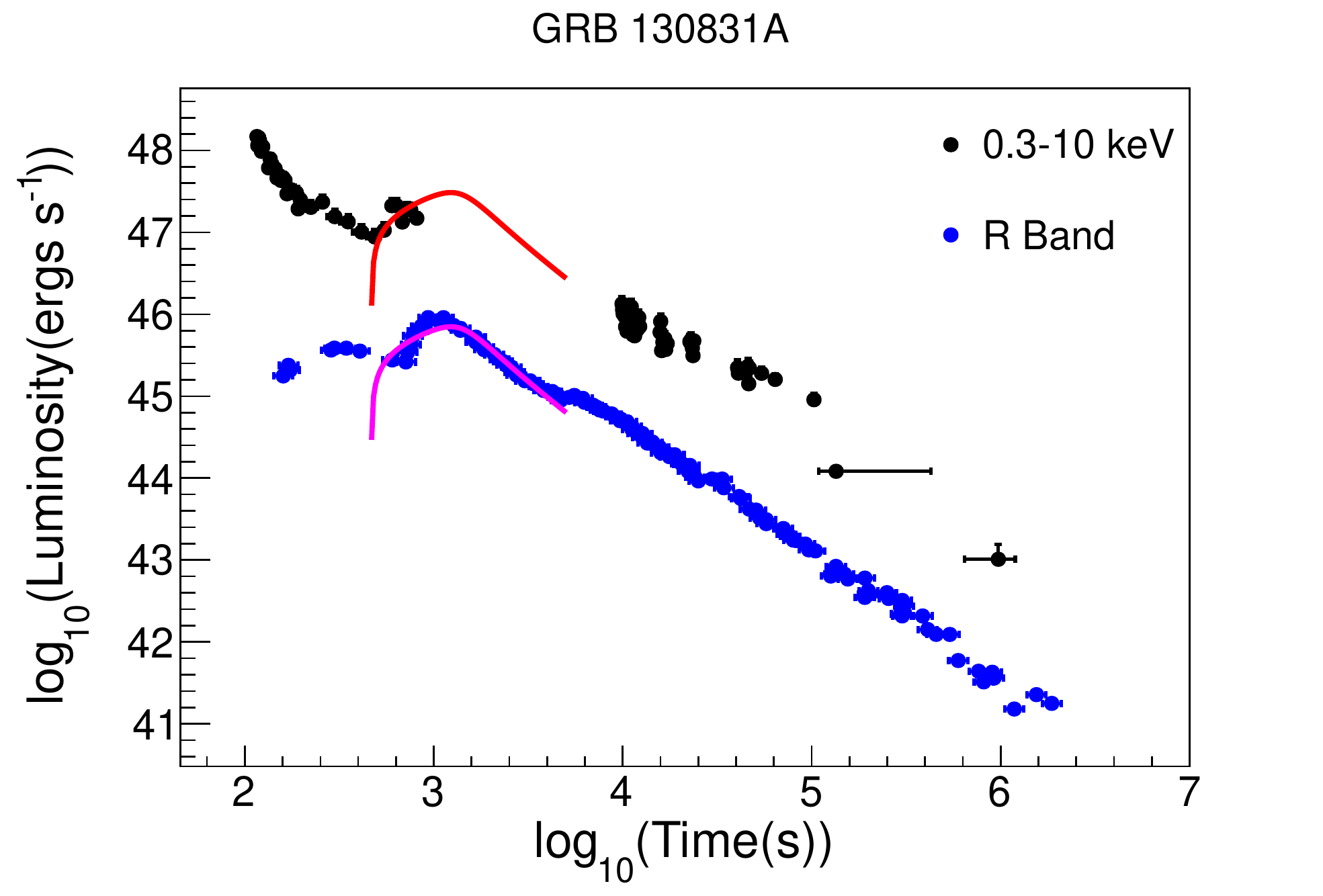}
        \includegraphics[width=5.86cm,height=6.5cm]{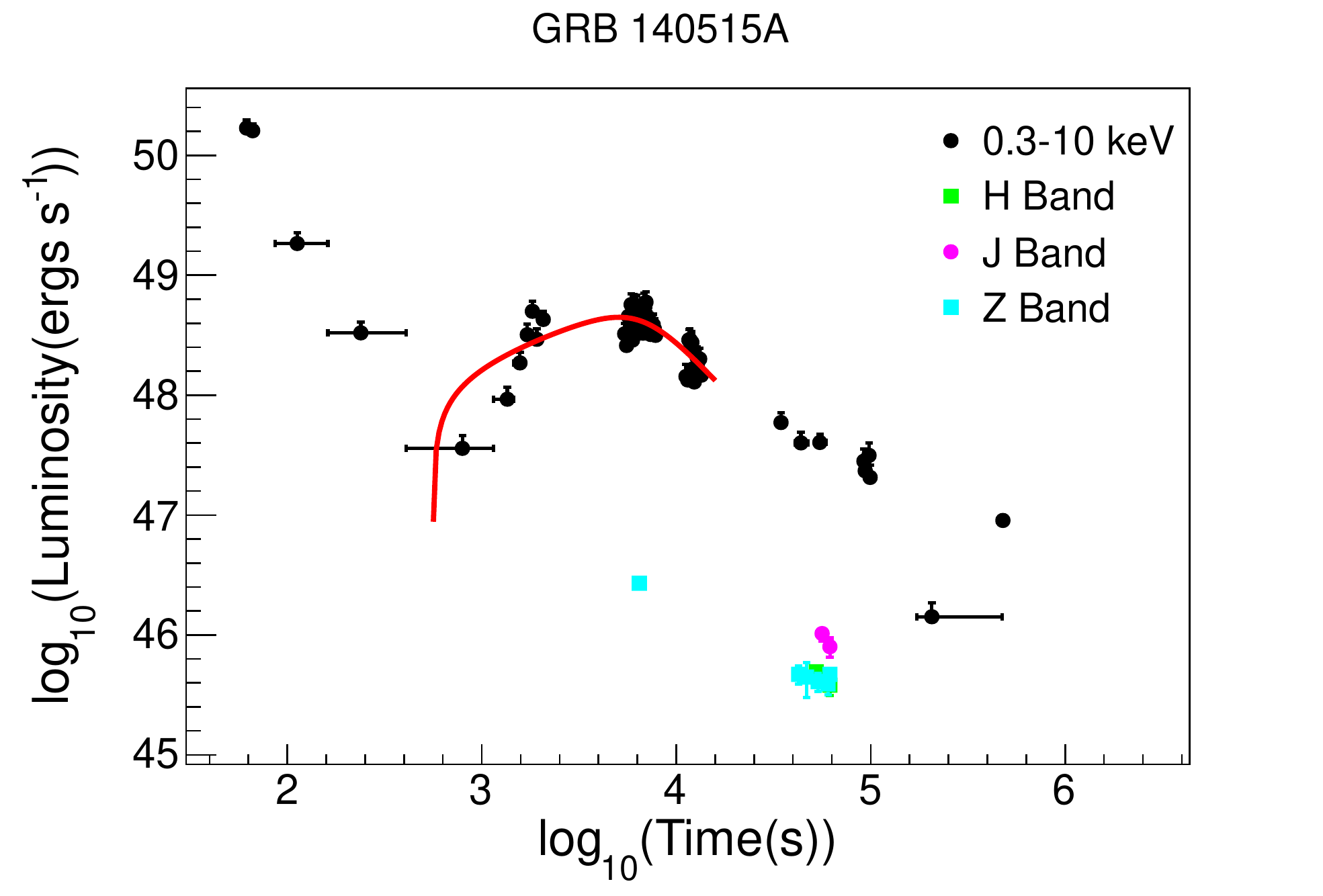}
        \includegraphics[width=5.86cm,height=6.5cm]{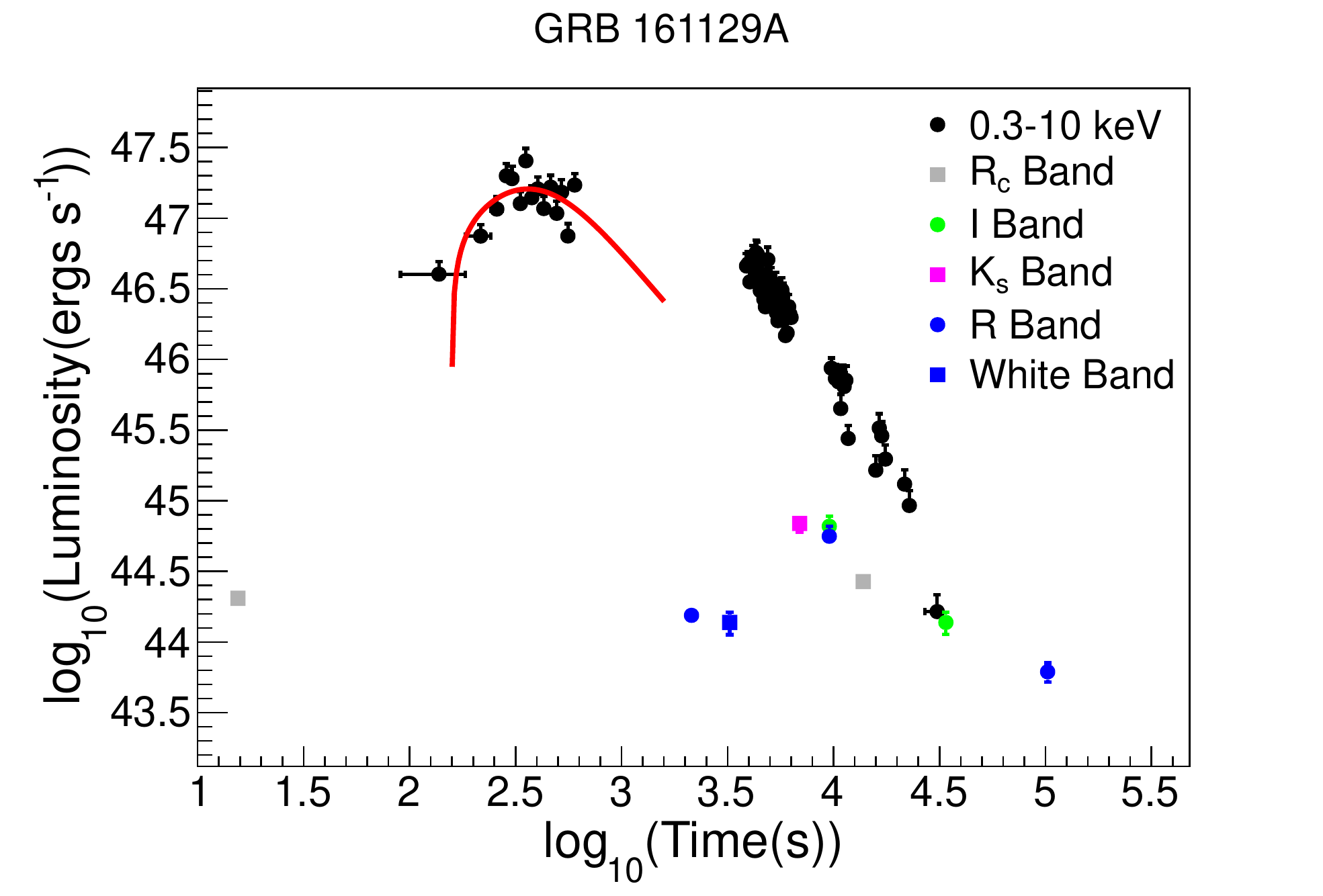}
        \includegraphics[width=5.86cm,height=6.5cm]{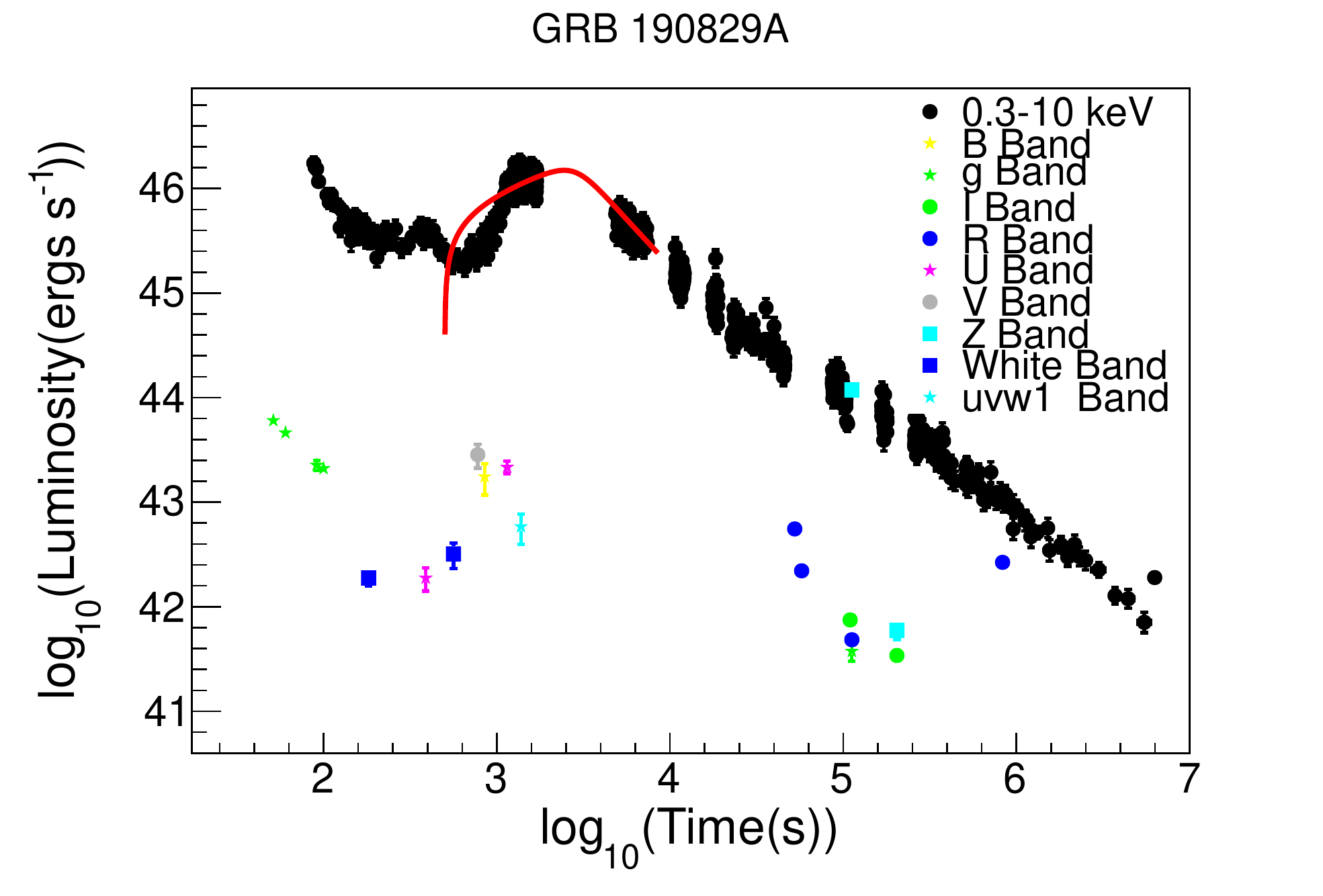}
        \caption{-Continued}
        \label{fittingresult-2}
\end{figure}

 \begin{figure}[hbt]
        \figurenum{5}
	 \centering
	 \includegraphics[width=0.32\textwidth,height=6.5cm]{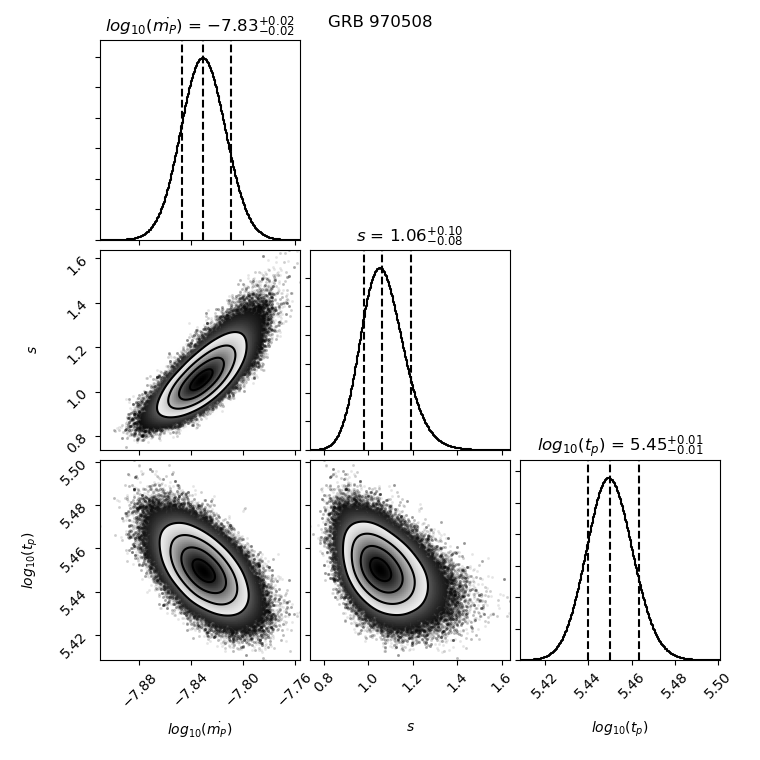}
	\includegraphics[width=0.32\textwidth,height=6.5cm]{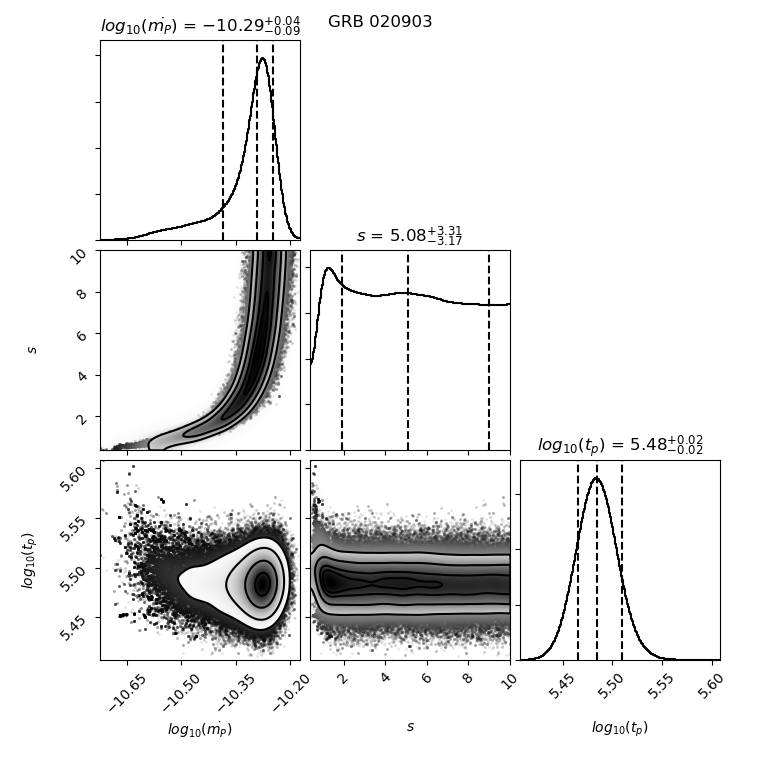}
       \includegraphics[width=0.32\textwidth,height=6.5cm]{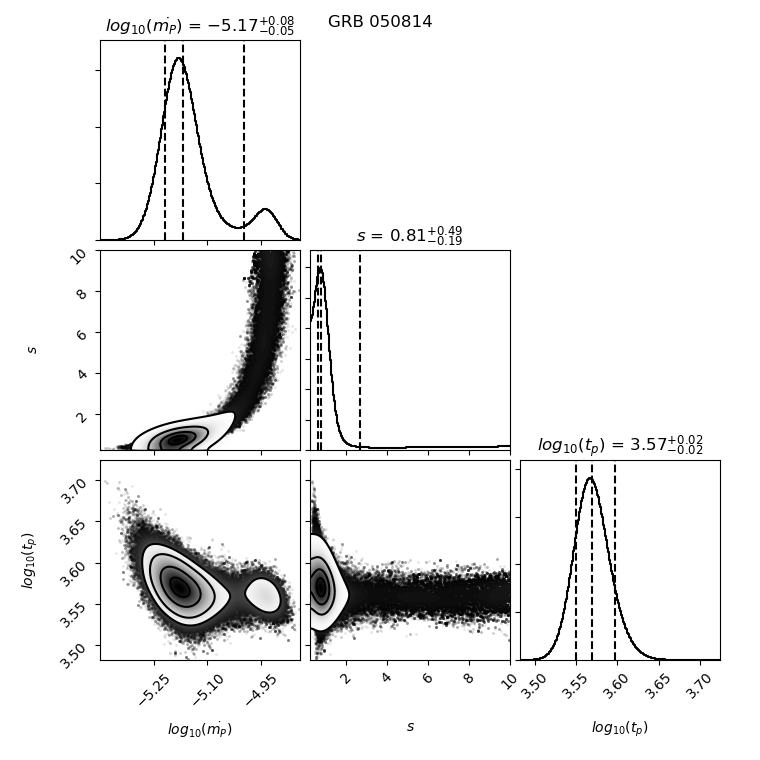}
       \includegraphics[width=0.32\textwidth,height=6.5cm]{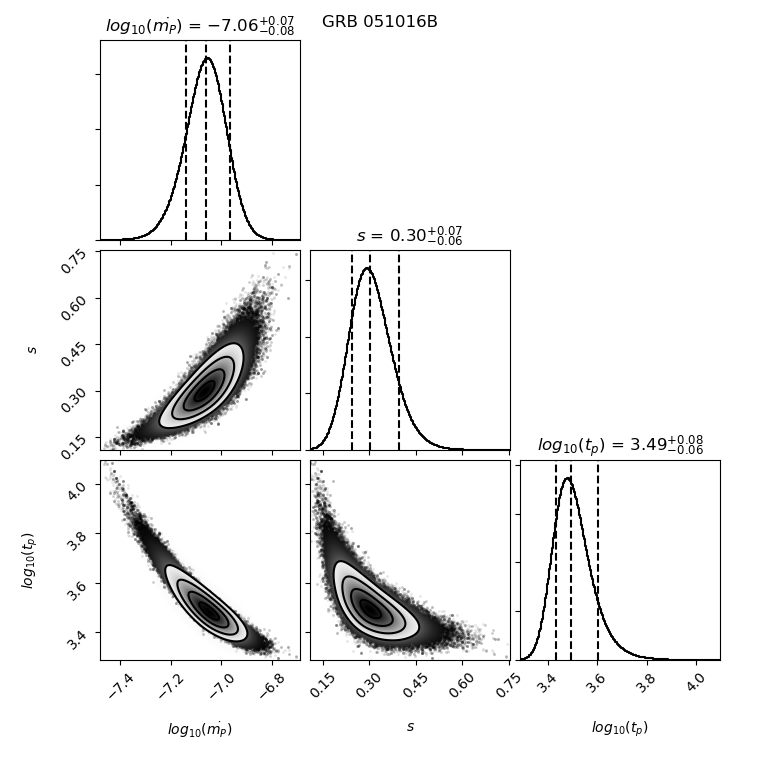}
       \includegraphics[width=0.32\textwidth,height=6.5cm]{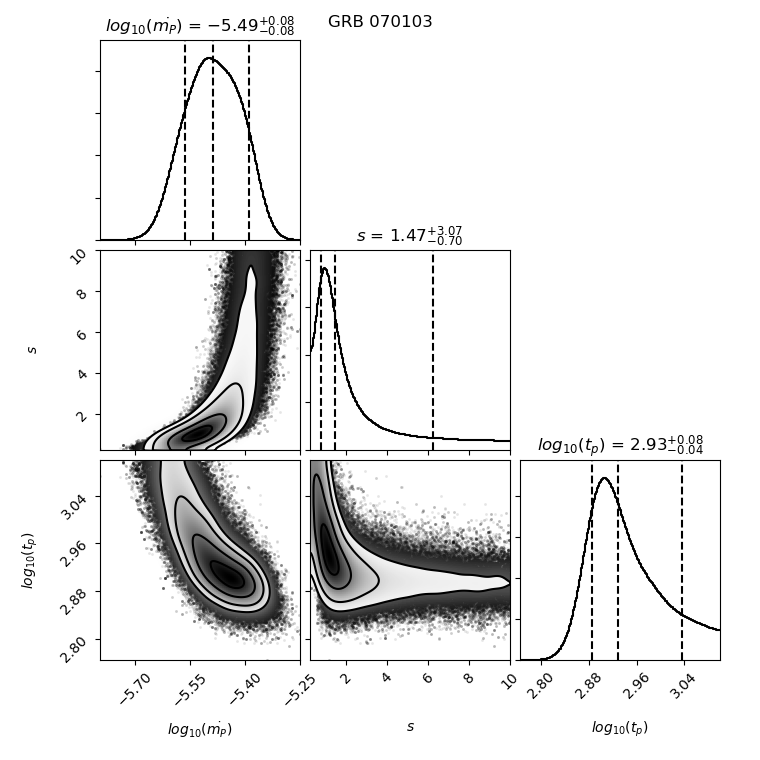}
       \includegraphics[width=0.32\textwidth,height=6.5cm]{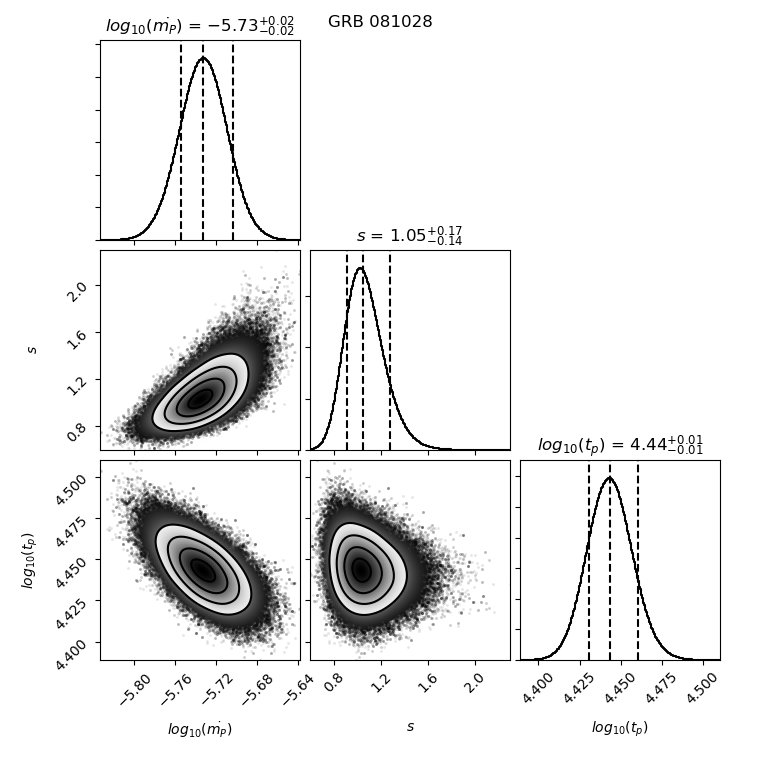}
       \includegraphics[width=0.32\textwidth,height=6.5cm]{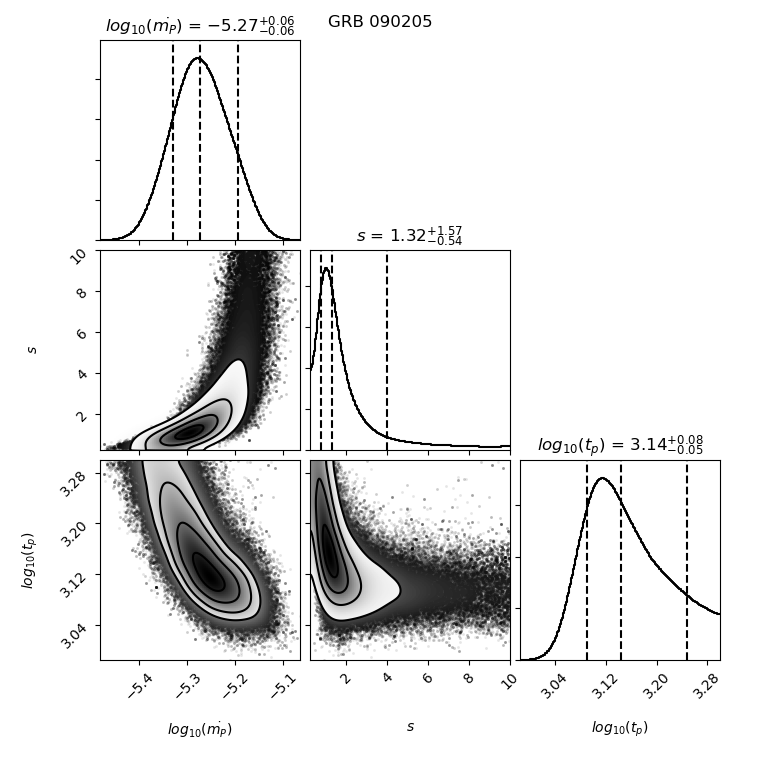}
       \includegraphics[width=0.32\textwidth,height=6.5cm]{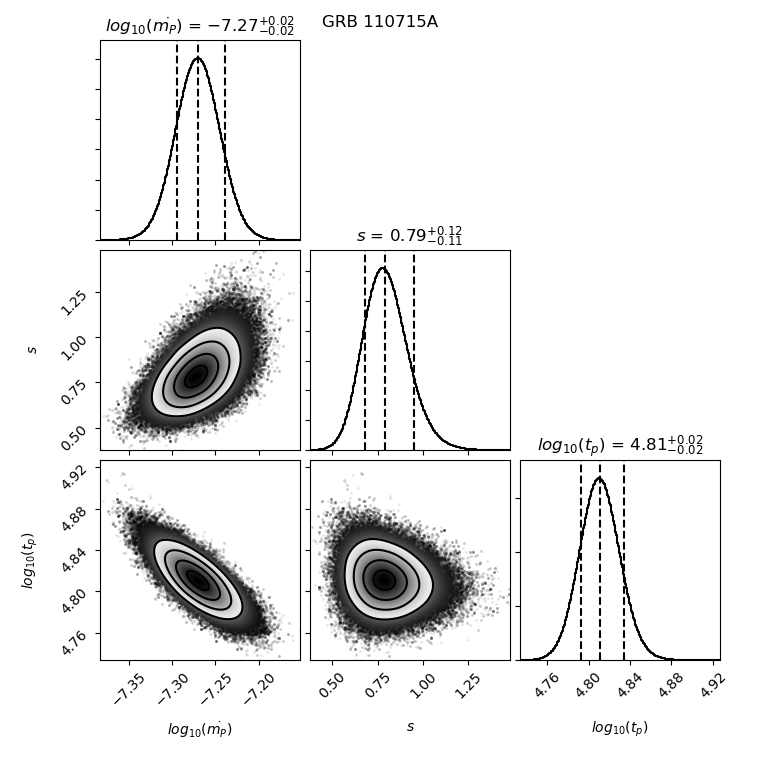}
       \includegraphics[width=0.32\textwidth,height=6.5cm]{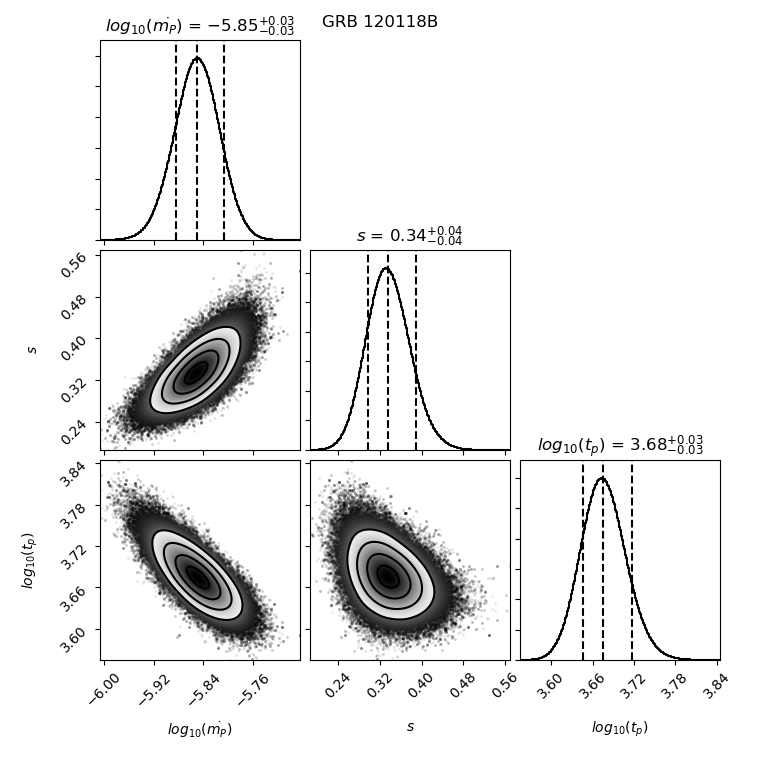}
       \caption{The corner plots  of the free parameters posterior probability distribution for the fitting  of the giant X-ray or optical bump in  the silver  sample.}
	\label{corner-2}
\end{figure}

 \begin{figure}[hbt]
       \figurenum{5}
	   \centering
       \includegraphics[width=0.32\textwidth,height=6.5cm]{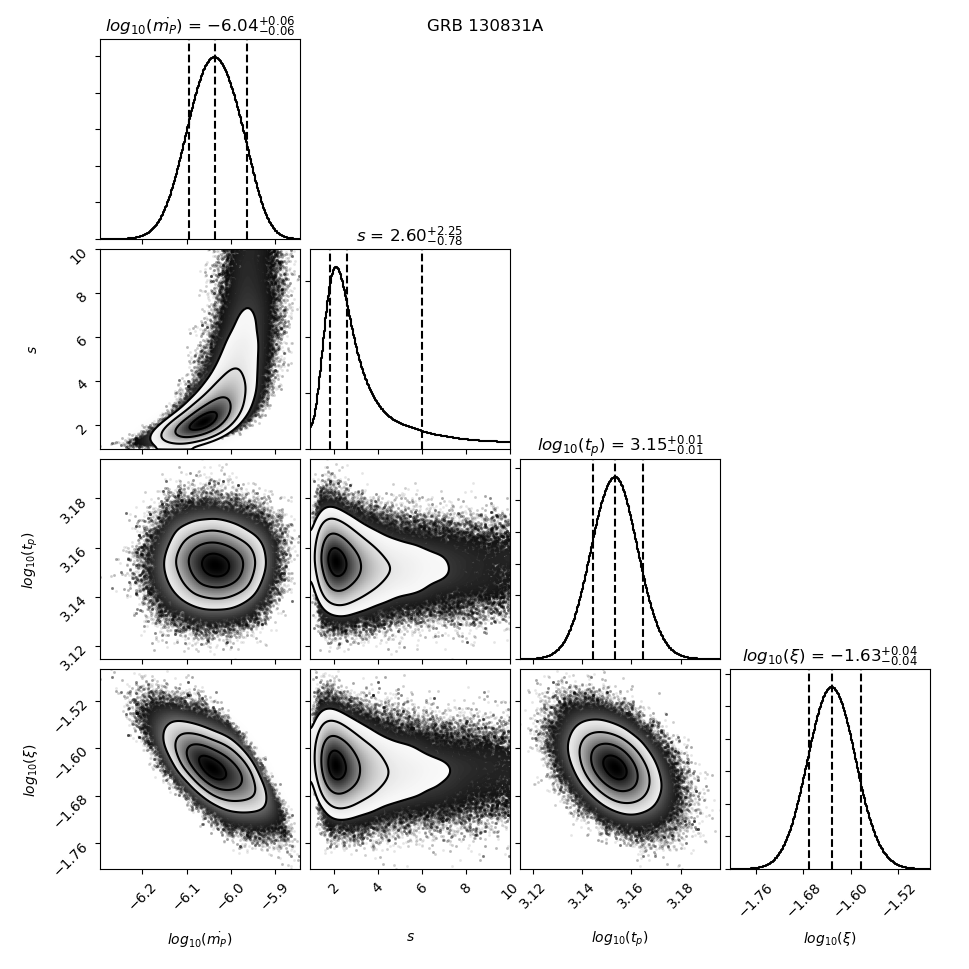}
       \includegraphics[width=0.32\textwidth,height=6.5cm]{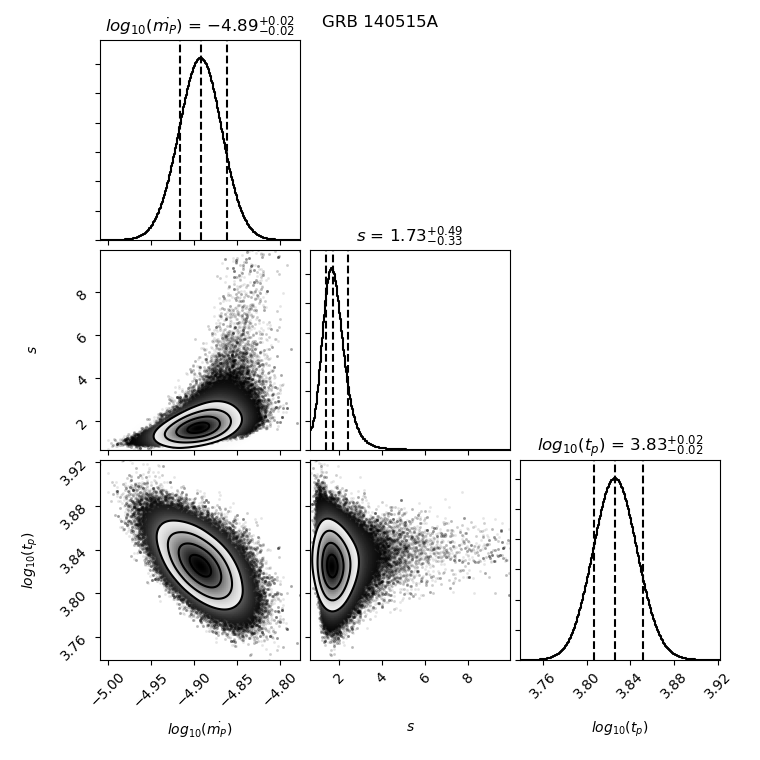}
       \includegraphics[width=0.32\textwidth,height=6.5cm]{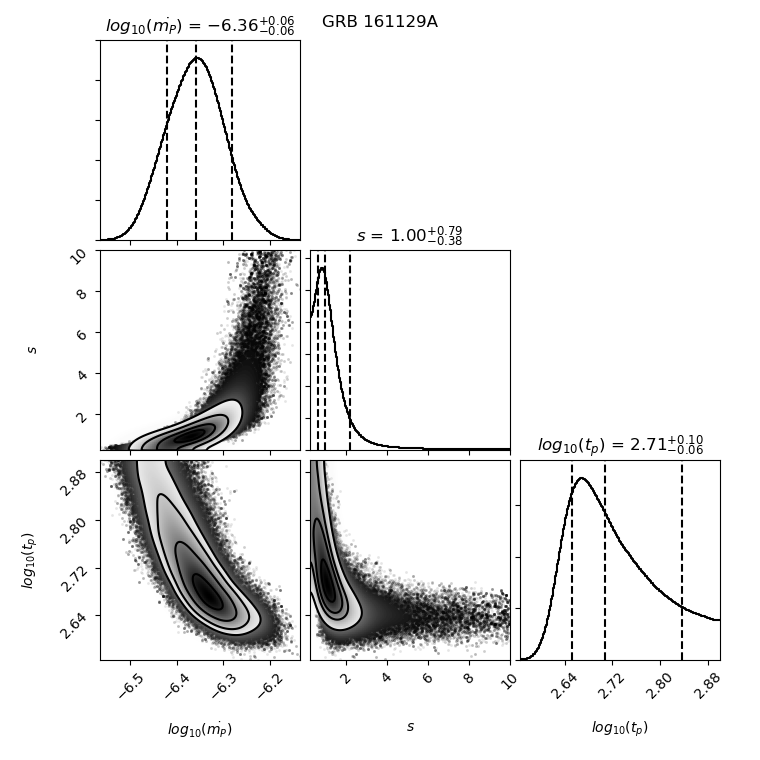}
       \includegraphics[width=0.32\textwidth,height=6.5cm]{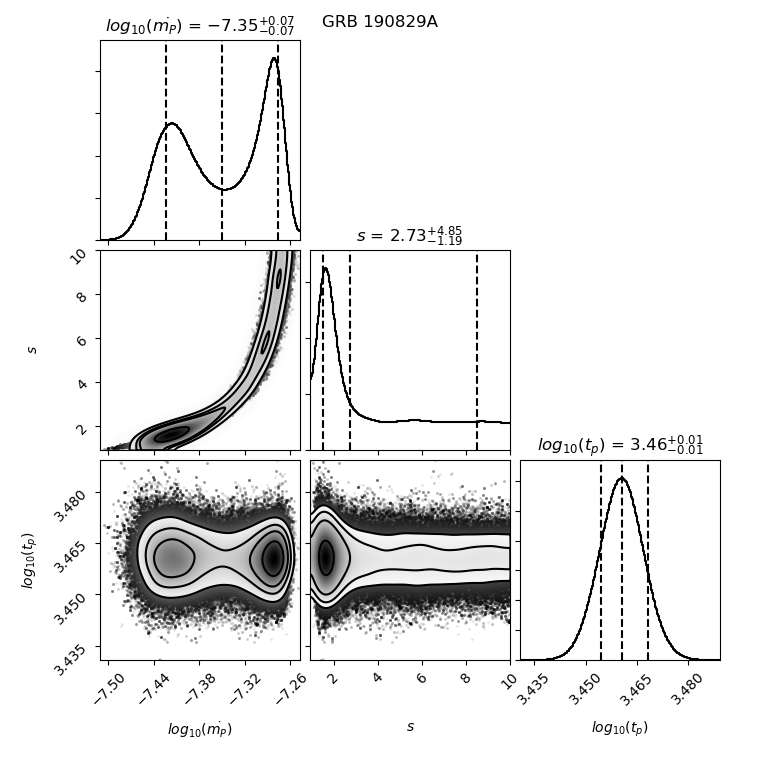}
       \caption{-Continued}
       \label{corner-2}
\end{figure}

\setlength{\tabcolsep}{1mm}
\begin{deluxetable*}{ccCclccccc}[b!]
\tablecaption{MCMC multiband fitting results for the gold sample \label{table-2}}
\tablecolumns{6}
\tablenum{2}
\tablewidth{0pt}
\tablehead{
\colhead{GRBname} &
\colhead{$\log_{10}(\dot{m}_{p})$}&
\colhead{$s $} &
\colhead {$\log_{10}(t_{\rm p})$}&
\colhead {$\log_{10}(\xi)$}&
\colhead {$M_{\rm acc}$}&
\colhead{$B_{P,15}$}&
 \colhead{$r_{p}$}&
\colhead{$\chi^{2}$}\\
\colhead{} & \colhead{} & \colhead{}  &\colhead{s} & \colhead{} & \colhead{$10^{-5} M_{\odot}$}&\colhead{$10^{-5}$}&\colhead{\rm$10^{10}$ cm}&
}
\startdata
060206  & $ -5.12\pm0.017$ & $ 0.53\pm0.01$&$3.91\pm0.003$ &$-0.71\pm0.016 $& 11723.21$\pm$ 2311.23 &$7402.21\pm151.31$ &$ 10.76\pm0.71 $&310.77\\
060906&$ -6.01\pm0.03$ & $ 1.05^{+0.14}_{-0.12}$ & $ 4.14\pm0.01$&$-0.75\pm0.02$ &$ 2020.61 \pm310.32 $& 2665.05 $\pm$621.32 &16.05$\pm$0.25&49.58\\
071010A& $ -7.74\pm0.04$ & $ 0.76^{+0.09}_{-0.07}$&$4.84\pm0.01$ &$-0.68\pm0.03 $& 214.52 $\pm$61.32 & 350.78$\pm$52.23&$ 119.13\pm3.45 $&41.87\\
081029& $ -5.46\pm0.02$ & $ 0.67\pm0.05$&$4.01\pm0.01$ &$-0.62\pm0.02 $& 6903.02 $\pm$482.32 &4900.13$\pm$320.32&$ 12.78\pm0.34 $&327.28\\
100418A& $ -8.38\pm0.03$ & $ 0.22\pm0.01$&$4.96\pm0.02$ &$-1.26\pm0.02 $& 112.91$\pm$37.32 &150.45$\pm$14.24&$ 184.52\pm5.71 $&112.31\\
100901A& $ -6.5\pm{0.02}$ & $ 0.98\pm0.07$&$4.63\pm0.01$ &$-0.63^{+0.01}_{-0.02} $& 2810.32$\pm$212.43 &1431.12$\pm$643.31&$ 64.71\pm1.23 $&256.94\\
\enddata
\end{deluxetable*}

 \setlength{\tabcolsep}{1mm}
 \begin{deluxetable*}{ccCclcccc}[b!]
\tablecaption{MCMC fitting results for the silver  sample \label{table-3}}
\tablecolumns{6}
\tablenum{3}
\tablewidth{0pt}
\tablehead{
\colhead{GRBname} &
\colhead{$ \log_{10}(\dot{m}_{p})$}&
\colhead{$s $} &
\colhead {$\log_{10}(t_{p})$}&
\colhead {$M_{\rm acc}$}&
\colhead{$B_{P,15}$}&
 \colhead{$r_{p}$}&
\colhead{$\chi^{2}$}\\
\colhead{}  & \colhead{} &\colhead{} & \colhead{s} & \colhead{$10^{-5} M_{\odot}$}&\colhead{$10^{-5}$ }&\colhead{\rm $10 ^{10}$cm}&
}
\startdata
970508  & $ -7.83\pm0.02$ & $ 1.06^{+0.1}_{-0.08}$&$5.45\pm0.01$ & 763.32$\pm$15.34 &311.13$\pm$13.13&$ 303.51\pm17.42 $&157.54\\
020903 &  $ -10.29^{+0.04}_{-0.09}$ & $ 5.08^{+3.31}_{-3.17}$&$5.48\pm0.02$ & 2.22$\pm$0.87 &18.27$\pm$9.13&$ 766.67\pm49.01 $ &11.29\\
050814& $ -5.17^{+0.08}_{-0.05}$ & $ 0.81^{+0.49}_{-0.19}$&$3.57\pm0.02$ & 4938 .12$\pm$201.53 & 6684.72 $\pm$255.08&$ 5.26\pm0.16$ &  35.66\\
051016B & $ -7.06^{+0.07}_{-0.08}$ & $ 0.30^{+0.07}_{-0.06}$&$3.49^{+0.08}_{-0.06}$ & $84.62 \pm11.21$&  890.27$\pm$152.17&$14.76\pm1.67$&13.94 \\
070103 & $ -5.49\pm0.08$ & $ 1.47^{+3.07}_{-0.70}$&$2.93^{+0.08}_{-0.04}$ & $386.21\pm 60.31$ & 4435.12$\pm$1135.76&$3.15\pm0.31$ &7.22 \\
$081028^{a}$&-5.73$\pm$0.02&$1.05^{+0.17}_{-0.14}$&4.44$\pm$0.01&6803.71$\pm$112.58& 3451.85$\pm$732.11&28.95$\pm$0.45&150.32\\
090205 & $ -5.27\pm0.06$ & $ 1.32^{+1.57}_{-0.54}$&$3.14^{+0.08}_{-0.05}$ & $  1116.32\pm$142.32 & 5748.34$\pm$ 1732.23&$ 2.96\pm0.31$&6.13\\
110715A & $ -7.27\pm0.02$ & $ 0.79^{+0.12}_{-0.11}$&$4.81\pm0.02$ & $  640.24\pm$31.21 & 612.23$\pm$25.22&$ 113.34\pm5.21$&72.48\\
$111209A^{b}$&-3.7&1.9&3.31& 21045.93& 23743.83&1.21&752.47\\
120118B & $ -5.85\pm0.03$ & $ 0.34\pm0.04$&$3.68\pm0.03$ &2873.32$\pm$691.32 &3481.42$\pm$642.13 &$9.07\pm0.51 $&27.36\\
$121027A^{c}$&-3.21&1.9&3.47&90000&5958&9.36&290.34\\
130831A &  $ -6.04\pm0.06$ & $ 2.6^{+2.75}_{-0.78}$&3.15$\pm$0.01& 160.01$\pm$68.51&2302.35$\pm$1123.31 &$64.71\pm1.23 $&39.76\\
140515A  & $ -4.89\pm0.02$ & $ 1.73^{+0.49}_{-0.33}$&$3.83^\pm0.02$ &13357.62 $\pm$1131.32 &6431.23$\pm$ 732.17 &$ 6.97\pm0.23$&51.32\\
161129A& $ -6.36\pm0.06$ & $ 1.00^{+0.79}_{-0.38}$&$2.71^{+0.1}_{-0.06}$ & $33.23\pm1.91 $&1224.51$\pm$230.02  &$  5.72\pm0.74$ &18.42\\
190829A&$-7.35\pm0.07$ & $ 2.73^{+4.85}_{-1.19}$&$3.46\pm0.01$ & 11.02$\pm$5.23 &512.21$\pm$44.14&$  71.72\pm5.31$&152.73\\
\enddata
\tablenotetext{a}{Same as the gold sample, we also take $\xi$ as the free parameter for fitting GRB130831A and the fitting result is $\log10(\xi)=-1.63\pm0.04$. }
\tablenotetext{b}{The giant X-ray bump  fitting results of GRB 111209A are adopted from \cite{yu15}. }
\tablenotetext{c}{The giant X-ray bump  fitting results of GRB 121027A are adopted from \cite{wu13}. }
\end{deluxetable*}

\section{Discussion and Conclusion}\label{conclusion}

The giant X-ray bumps discovered in the afterglow of GRB 121027A and 111209A, have been proposed as the direct evidence to support that the late
central engine activity of long GRBs is likely due to the fall-back accretion process. In this work, we systemically searched all long GRBs detected between 1997 January and 2019 October, and found another 19 candidate GRBs showing giant X-ray  or optical bumps  in their afterglows. We have applied the fall-back accretion model to interpret the X-ray and optical bump data for the whole sample. We summarize our results as follows:

\begin{itemize}
\item We find that the X-ray and optical bump data for the gold sample and silver sample could be well interpreted by the fall-back accretion model within a fairly flexible and reasonable parameter space. For the six GRBs in the gold sample showing a simultaneous bump signature in both X-ray and optical observations, the X-ray and optical data could be well fitted simultaneously with $\xi = \eta_O/\eta_X \sim 0.1$, which happens to be consistent with the standard $F_{\nu}\propto\nu^{1/3}$ synchrotron spectrum below $E_p$. 

\item The fall-back accretion rate reaches its peak value $\sim 10^{-11}-10^{-4}M_{\odot} \ \text{s} ^{-1}$ at time $\sim10^{2}- 10^{5}~\rm s$, the constraints set by the progenitor's mass and metallicity are not strong, which makes the mass supply rate of the progenitor envelope material at late time fulfill the fall-back accretion rate requirement. The lower limit of total accretion mass could be as low as $\sim 10^{-5} M_{\odot}$, which means even with a very small fraction of the progenitor's envelop falling back, it is still possible to generate the giant X-ray bump and optical feature.
\item The typical fall-back radius is around $10^{10}-10^{12}$ cm, which is consistent with the typical radius of a Wolf-Rayet star.
\item The  jet energy from the fall-back accretion is  linearly correlated with  the isotropic-equivalent radiation energies of GRB prompt phase in the  $1-10^{4}$ keV band, implying that the fall-back accretion is correlated with the prompt phase accretion.
\end{itemize}

In conclusion, our results provide additional support for core collapse from Wolf-Rayet star as the progenitors of long GRBs, whose late  central engine activity is very likely caused by the fall-back accretion process.

There are two possible reasons why most long GRBs do not show a giant X-ray and optical bump: firstly, if the progenitor of long GRBs produce a high-energy supernova when its core collapses into a BH, the supernova shock may eject most of envelope materials and leave too little  material falling back into the BH; secondly, during the fall-back accretion progress, most of the jet energy injects into the GRB blast wave, while the energy that undergoes internal dissipation is weak.

It is worth noting that in Figure \ref{mdot},  we  make a direct comparison between the required peak accretion rate ($\dot{M}_{\rm acc}$) and the fall-back mass rate ($\dot{M}_{\rm fall}$). If considering that a good fraction of fall-back mass could be taken away by the accretion disk outflow, $\dot{M}_{\rm fall}>\dot{M}_{\rm acc}$ is required, so that low metallicity progenitor stars would become more favored. On the other hand, when calculating the fall-back mass rate, we did not consider the angular momentum distribution of the progenitor, which is approximately valid for a slowly rotating progenitor. In the future, detailed studies for the evolution model of the mass supply rate for the progenitors with different mass, metallicity and rotation speed would help us to better constrain the progenitor properties of long GRBs.

\acknowledgments
We thank the anonymous referee for the helpful comments that have helped us to improve the presentation of the paper.  This work is supported by the National Natural Science Foundation of China (NSFC) under Grant No. 11722324,11690024,11633001,11773010 and U1931203, the Strategic Priority Research Program of the Chinese Academy of Sciences, Grant No. XDB23040100 and the Fundamental Research Funds for the Central Universities. LDL is supported by  the National Postdoctoral Program for Innovative Talents (Grant No. BX20190044), China Postdoctoral Science Foundation (Grant No. 2019M660515)  and ``LiYun'' postdoctoral fellow of Beijing Normal University.
\software{XSPEC(\cite{arnaud96}), HEAsoft(v6.12;\cite{heasarc14}), root(v5.34;\cite{brunRademakers97}), emcee(v3.0rc2;\cite{formanmackey13}),corner(v2.0.1;\cite{foremanmackey16})    }

\end{document}